\documentclass[11pt,twocolumn,aps,prb,reprint]{revtex4-1}

\usepackage{braket}

\usepackage{graphicx,amsmath,amssymb,braket,bm,color}

\begin{document}

\title{Plasmon-assisted resonant tunneling in graphene-based heterostructures}
\author{V. Enaldiev, A. Bylinkin, D. Svintsov}
\affiliation{Laboratory of 2d Materials' Optoelectronics, Moscow Institute of Physics and Technology, Dolgoprudny 141700, Russia}
 
\begin{abstract}
We develop a theory of electron tunneling accompanied by carrier-carrier scattering in graphene - insulator - graphene heterostructures. Due to the dynamic screening of Coulomb interaction, the scattering-aided tunneling is resonantly enhanced if the transferred energy and momentum correspond to those of surface plasmons. We reveal the possible experimental manifestations of such {\it plasmon-assisted} tunneling in current-voltage curves and plasmon emission spectra of graphene-based tunnel junctions. We find that inelastic current and plasmon emission rates have sharp peaks at voltages providing equal energies, momenta and group velocities of plasmons and interlayer single-particle excitations. The strength of this resonance, which we call {\it plasmaronic resonance}, is limited by interlayer twist and plasmon lifetime. The onset of plasmon-assisted tunneling can be also marked by a cusp in the junction $I(V)$-curve at low temperatures, and the threshold voltage for such tunneling weakly depends on carrier density and persists in the presence of interlayer twist.
\end{abstract}

\maketitle

\section{Introduction}

The plasmons supported by two-dimensional (2d) electron systems, including graphene, can be confined at the distance by two orders of magnitude smaller than free-space light wavelength~\cite{Koppens_nano_imaging_hBN}. High field confinement in graphene plasmon modes leads to an enhanced electron-plasmon interaction~\cite{Graphene_plasmonics-2} and a number of interesting phenomena. The formation of new quasiparticles of bound holes and plasmons, called plasmarons~\cite{Bostwick}; perfect light absorption~\cite{Perfect_absorption}; ultrafast recombination of photoexcited carriers mediated by plasmon emission~\cite{Rana_IEEE,Watanabe_gain_enhancement} are among the brightest manifestations of electron-plasmon coupling. The strength of light-matter interaction is the highest for acoustic plasmon modes supported by 2d bilayers and gated 2d systems~\cite{Ryzhii-plasmons, Koppens_AcousticPlasmons}; thereat the electromagnetic energy is concentrated at the length scale of interlayer spacing $d$. Such bilayers are actively investigated both as building blocks of novel resonant tunneling diodes~\cite{Novoselov_APL_RTD}, transistors~\cite{britnell2013resonant}, and as a polygon for the fundamental studies of tunneling in the presence of chirality~\cite{Chiral_tunneling} and field-controlled interlayer twist~\cite{Novoselov_Magnetic_tunneling}.

What new effects can stem from strong electron-plasmon interaction in tunnel-coupled graphene layers? One might naively expect the emergence of steps in the dependence of tunnel conductivity on interlayer bias~\cite{Phonon_assisted_graphene}. These steps {\it commonly} occur at voltages $V = \hbar \omega_i/e$~\cite{Esaki_Phonon}, where $\hbar \omega_i$ are the characteristic energies of collective excitations. However, the plasmon dispersion is soft in two dimensions, i.e. its frequency tends to zero in the long-wavelength limit. Therefore, generally there is no preferable frequency for plasmon-assisted tunneling in 2d, and plasmonic fingerprints in tunneling can appear only at very specific conditions~\cite{Plasmon-assisted_Eaves,Belenov_Emission_resonant}, e.g., at the anticrossing of different plasmon modes~\cite{Kempa_PlasmonAssisted}. In this paper, we find that in aligned graphene double layers the plasmon-assisted tunneling can surprisingly lead to a resonant enhancement of tunnel current. The interlayer bias providing the resonance corresponds to the energy, momentum, and group velocity matching between plasmons and interlayer single-particle excitations. This effect is inherent to the graphene's linear band structure. In some sense, it is similar to the formation of plasmarons a single graphene layer due to the consonance between plasmon and electron motion~\cite{Bostwick} -- for this reason, we call this effect {\it plasmaronic resonance}.

The plasmaronic resonance can manifest itself also in the spectra of the tunnel junctions' electroluminescence~\cite{Lambe_PRL,Parzefall2017}. We show that in aligned layers the integrated luminescence can demonstrate a strong spike at the resonant voltage, and be considerably larger than the recently observed luminescence from twisted layers~\cite{Yadav_THz}. In the presence of twist, an only remainder of plasmons is the fine structure of low-temperature $I(V)$-curves due to switch-on of plasmon-assisted tunneling.

\begin{figure}
    \center{\includegraphics[width=0.8\linewidth]{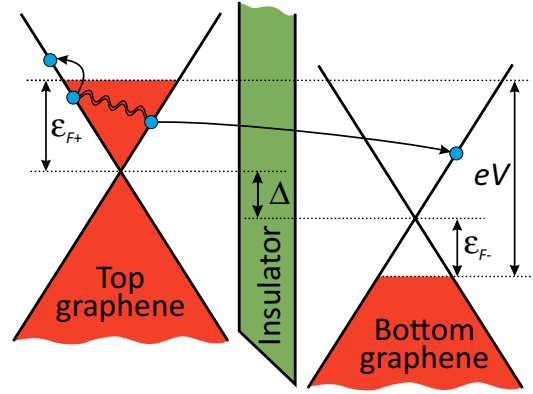} }
    \caption{Band diagram of a graphene--hBN--graphene tunnel junction and a schematic view of electron tunneling accompanied by scattering. $eV$ is the applied voltage, $\Delta$ is the interlayer band offset, $\varepsilon_{F\pm}$ ate the Fermi energies in top ($+$) and bottom ($-$) layers}
    \label{fig:band_diagram}
\end{figure}

So far, a considerable work has been done in the studies of plasmon spectra and damping/gain in graphene double layers~\cite{Hwang_PRB_2GL,Voltage_controlled}, including the tunneling effects~\cite{Plasmons_Coupled, Berardi_APL}. On the other hand, there exists a number of theoretical works on various tunneling channels in these systems: elastic~\cite{Jena-single-particle,Vasko_PRB,Brey_PRA}, including the effects of finite carrier lifetime~\cite{Polini_tunneling_lifetime}, phonon-\cite{Phonon-assisted-theory} and photon-assisted~\cite{Ryzhii_DGL_laser}. However, no attempt has been made to calculate the plasmon-assisted tunnel current. The present theory is based on the calculation of electron interlayer transition rates due to the dynamically screened Coulomb interaction with other carriers. By extracting the plasmon-pole contribution from the scattering-assisted tunnel current, we are able to find the current due to plasmons, and plasmon emission rates as well. As an added benefit, our formalism allows to calculate the full inelastic tunnel current, including the non-resonant scattering contributions. However, as we find, the largest fraction of Coulomb-scattering-assisted current is due to emission of plasmons. We also present a simplified approach for plasmon-assisted current based on the calculation of electron tunneling due to fluctuating electric fields, and evaluation of the magnitude of fluctuations with second quantization procedure. These two approaches coincide in the limit of weak electromagnetic dissipation.

\section{Theory of many-particle and plasmon-assisted tunneling}

Electron states in coupled graphene layers with small interlayer twist can be labelled by in-plane momentum ${\bf p}$, the band index $s=\pm 1$ and the index $l=\pm 1$ governing the vertical localization of electron wave function~\cite{Vasko_PRB}. The respective energies are $\epsilon^{ls}_{\bf p} = s p v_0 + l \sqrt{\Delta^2/4 +\Omega^2}$, where $v_0$ is the Fermi velocity, $\Delta$ is the voltage-induced splitting of bands in neighboring layers (band offset, see Fig. 1), and $\Omega$ is the tunnel splitting. For strong bias and/or weak tunneling,  $\Delta \gg \Omega$, the state $l=+1$ can be regarded as belonging to the top layer and $l=-1$ -- to the bottom one. The Coulomb interaction couples the states in neighboring layers and induces inelastic tunneling current. The current from the forward biased (top) layer can be presented as
\begin{equation}
\label{Inelastic}
I_{t\rightarrow b } = e g^2 \sum\limits_{
\begin{smallmatrix} 
 {\bf p}{\bf p_1}{\bf q} \\ 
 ls{s_1}s'{s'_1} 
\end{smallmatrix}
} W_{fi} 
 f^{+s}_{\bf {p}} [1 - f^{-s'}_{\bf p'} ] f^{ls_1}_{{\bf p}_1} [1 - f^{ls'_1}_{{\bf p}'_1}],
\end{equation}
where $g=4$ is the spin-valley degeneracy factor, $f^{ls}_{\bf p}$ is the occupancy of the quantum state with respective set of quantum numbers, and $W_{fi}$ is the golden rule transition probability between two-particle states $\ket{+ {\bf p} s, l {\bf p}_1 s_1}$ and $\ket{- {\bf p}' s', l' {\bf p'}_1 s'_1}$ due to Coulomb interaction. The full current including the reverse component is $I = I_{t\rightarrow b} [ 1 - e^{-eV/kT} ]$,
where $V$ is the interlayer voltage. We note that $eV = \Delta + \varepsilon_{F+} - \varepsilon_{F-}$, where $\varepsilon_{F\pm}$ are the Fermi energies in respective layers. The quantities $\Delta$ and $\varepsilon_{F\pm}$ can be controlled independently with doping or extra gates.

A sequence of transformations common in the theory of Coulomb scattering phenomena~\cite{Zheng_Lifetime} leads us to a particularly simple form of the tunneling current (we set $\hbar \equiv 1$)
\begin{multline}
\label{Many-particle}
I_{t \rightarrow b}=\frac{2e}{\pi} \int\limits_{-\infty }^{+\infty }{d\omega }\sum\limits_{{\bf q},l} \Pi''_{+-}( q, \omega ) \Pi''_{ll}\left( q,\omega  \right)\times \\
\left| V_{+l,-l} \right|^2 N_{\omega -eV} \left[ N_\omega + 1 \right],
\end{multline}
where $N_\omega = [e^{\omega/T} - 1]^{-1}$ is the Bose distribution and $\Pi_{ll'} = \Pi'_{ll'} + i \Pi''_{ll'} $ is the intra- ($l=l'$) or inter-layer ($l\neq l'$) polarizability:
\begin{equation}
    \Pi_{ll'} = \frac{g}{A} 
    \sum\limits_{
    \begin{smallmatrix}
     {\bf p} ss'\\
     {\bf p}'={\bf p}+{\bf q}
    \end{smallmatrix}
    }{\frac{1+ss' \cos\theta_{\bf pp'}}{2}\frac{ f^{ls}_{{\bf p}} -f^{l's'}_{{\bf p}'}}{\omega + i \delta - \epsilon^{ls}_{\bf p} + \epsilon^{l's'}_{{\bf p}'}}}.
\end{equation}
The amplitude $ V_{+l,-l} $ describes the tunneling of electron from top to bottom layer upon Coulomb interaction with a carrier in the $l$-th layer. The latter depends on frequency and wave vector, primarily, due to the dynamic screening of Coulomb interaction. In the dipole approximation, the transition amplitude becomes 
\begin{multline}
\label{amplitudes}
    V_{+l,-l} = \frac{ V_0 ({\bf q}) }{ \varepsilon({\bf q},\omega) } \frac{z_\pm}{d}(1-e^{-q d})\times \\
    \left[ 1 - V_0({\bf q}) \Pi_{-l -l} ({\bf q},\omega) (1+e^{-q d}) \right],
\end{multline}
where $z_\pm = \int_{-\infty}^{+\infty} dz \psi_+(z) z \psi_-(z)$ is the coordinate matrix element between initial and final states, $V_0({\bf q}) = 2\pi e^2/\kappa|{\bf q}|$ is the bare Coulomb interaction in two dimensions, $\kappa$ is the background dielectric constant, and $\varepsilon({\bf q},\omega)$ is the dynamic screening function of the double layer~\cite{Hwang_PRB_2GL}. The factor $1-e^{-q d}$ in Eq.~(\ref{amplitudes}) is due to the long range of Coulomb interaction. Indeed, the potential created by an electron is almost identical on both layers. Hence, the tunneling is weakened for long wavelengths (or slowly varying potentials) $q \rightarrow 0 $. This prefactor cancels the low-angle scattering singularity in bare Coulomb interaction, making the full expression for inelastic current convergent even in the absence of screening. Still, the screening has an important impact on interlayer tunneling, which is seen from the resonant enhancement of matrix element at $\varepsilon({\bf q},\omega) \rightarrow 0$. This is nothing but tunneling accompanied by emission of surface plasmons. 

To extract the plasmonic contribution, one can expand the dielectric function in the vicinity of plasmon poles $\varepsilon({\bf q},\omega) \approx \left[\partial \varepsilon' / \partial\omega\right] (\omega - \omega^p_{\bf q}) + i \varepsilon''$ (here $p=+1$ ($-1$) corresponds to optical (acoustic) modes, for details see Appendix B). Assuming the electromagnetic dissipation to be small, $|\varepsilon''/\varepsilon'| \ll 1$, we arrive at the expression for {\it plasmon-assisted} component of the net tunneling current $I^{\rm pl}_{t\rightarrow b}$. This can be conveniently split into the emission ($I^{\rm pl, em}_{t\rightarrow b}$) and absorption ($I^{\rm pl,abs}_{t\rightarrow b}$) contributions
 \begin{gather}
 \label{Plasmon-assisted}
     I^{\rm pl, em}_{t\rightarrow b} = I^{\rm pl,em}_{t\rightarrow b} + I^{\rm pl,abs}_{t\rightarrow b},\\
     I^{\rm pl,em}_{t\rightarrow b} = 2\pi e \sum\limits_{{\bf q}p} \left|\frac{ e {\varphi^{p}_{{\bf q} \pm}}}{2}\right|^2 {\Pi''_{tb}}( {\bf q}, \omega^{p}_{\bf q}) [ N_{\omega^{p}_{\bf q}} + 1 ] {N_{\omega^{p}_{\bf q}-eV}}.
 \end{gather}
The quantity $(e \varphi^p_{\bf q})_{\pm}$ can be viewed as a matrix element of electron interaction with zero-point field of plasmon:
\begin{multline}
\label{MatrixElements}
    \frac{ (e {\varphi^{p}_{\bf q})^2_{\pm}}}{2} = V_0(\bm{ q})\left|\frac{z_\pm}{d}\right|^2 \left(1 - e^{-qd}\right)^2\times \\ 
    \times\dfrac{ \left[1-V_0(\bm{q})\Pi'_{++}(\bm{q},\omega_q^{p})(1 + e^{-qd}) \right]^2}{\frac{\partial\varepsilon'}{\partial\omega_q^{p}}\left|1- V_0(\bm{q})\Pi'_{++}(\bm{q},\omega_q^{p})(1 - e^{-2qd})\right|}.
\end{multline}
The same result could be obtained by calculating the electron transition rates due to the interaction with zero-point and thermal longitudinal fluctuations of electromagnetic field. The magnitude of these fluctuations can be found from quantum-classical correspondence~\cite{Belenov_Emission_resonant}, i.e. by equating the classical energy of electromagnetic field in the dispersive medium to $N_\omega \hbar\omega$. This procedure is described in detail in Appendix A.
 
\section{Manifestations of plasmon-assisted tunneling}
With the general formalism of calculation developed, we start discussing the possible experimental manifestations of plasmon-assisted tunneling. We consider three such effects: (1) resonant enhancement of tunnel current due to group velocity coincidence between plasmons and inter-layer single particle excitations, which we call {\it plasmaronic resonance} (2) plasmonic electroluminescence of graphene tunnel junctions (3) fine structure of the low-temperature $I(V)$-curves due to onset of the plasmon emission.

\subsection{Plasmaronic resonance in tunnel current}

The imaginary part of interlayer polarizability $\Pi''_{\pm} ({\bf q},\omega)$ in aligned layers has a square-root singularity at the threshold of interlayer excitations, $\Pi''_{\pm} ({\bf q},\omega) \propto | (\Delta \pm q v)^2 - \omega^2 |^{-1/2}$. This singularity can be explained as resulting from prolonged interaction between electron and hole with collinear momenta and, hence, equal velocities. Similar singularities exist in the polarizability of a single graphene layer at the threshold of Landau damping\cite{Das_Sarma_Plasmons} $\Pi''_{ll} ({\bf q},\omega) \propto |  q^2 v^2 - \omega^2 |^{-1/2} $,  and were recently assessed experimentally~\cite{Nonlocal_GraphenePlasmons}. A pronounced effect of such ''collinear singularities'' is that the plasmon {\it phase} velocity always lies above the Fermi velocity, and the Landau damping of plasmons is absent~\cite{Ryzhii-plasmons}.

The {\it group} velocity of plasmons can be, however, equal to or below the Fermi velocity. This applies both to the modes supported by a single layer and graphene double layer as well, and is shown in Fig. 2 B. When the line of interlayer tunneling singularities $\omega = \Delta + q v $ approaches the tangent with plasmon dispersion, the plasmon-assisted tunneling current is resonantly enhanced. The resonant interlayer band offset $\Delta^*$ is determined from
\begin{gather}
    \Delta^* + q v =  \omega^{p}_q,\\
    \partial \omega^{p}_q/\partial q = v.
\end{gather}
In the vicinity of resonance the plasmon-assisted contribution grows as 
\begin{equation}
\label{Resonant_current}
   I^{\rm pl,em}_{t\rightarrow b} (\Delta) \approx I_0  \left. \ln\left|\dfrac{q^2 \partial^2\omega_q/\partial q^2 }{2(\Delta - \Delta^*)}\right| \right|_{q=q^*},
\end{equation}   
where the large logarithm is evaluated at $q= q^*$ which is the momentum of plasmons in resonance with interlayer excitations. The characteristic current in Eq.~(\ref{Resonant_current}) is
\begin{equation}
  I_0 =  \left|\frac{ e {\varphi^{p}_{{\bf q} \pm}}}{2}\right|^2  \left. \dfrac{eq \left(N_\omega + 1 \right) N_{\omega - eV}\widetilde{\Pi}_{\pm}(q)}{2\pi \hbar^2\sqrt{ q v \partial^2\omega_q/\partial q^2 }} 
  \right|_{\begin{smallmatrix}
            q=q^*\\
            \omega = \omega_{q^*}
            \end{smallmatrix}
            }
\end{equation}
where $\widetilde{\Pi}_{\pm}$ is non-singular part of $\Pi_{\pm}$ (i.e. $\Pi_{\pm}$ without the square-root singularity). The logarithmic growth of the current at the resonance is limited by plasmon damping. The latter was assumed to be infinitesimal in Eq.~(\ref{Plasmon-assisted}) but is automatically taken into account in many-particle formalism, Eq.~(\ref{Many-particle}), where both real and imaginary parts of dielectric function contribute to screening. It is possible to show that the damping-limited resonant value of current is, roughly
\begin{equation}
\label{Scattering_limit}
    I^{\rm pl,em}_{t\rightarrow b} (\Delta^*) \approx 
    I_0 \left. \ln\left|\dfrac{q^2 \partial^2\omega_q/\partial q^2 }{\varepsilon'' / [\partial\varepsilon'/\partial\omega]}\right| \right|_{q=q^*}.
\end{equation}

\begin{figure}
   \center{\includegraphics[width=1.0\linewidth]{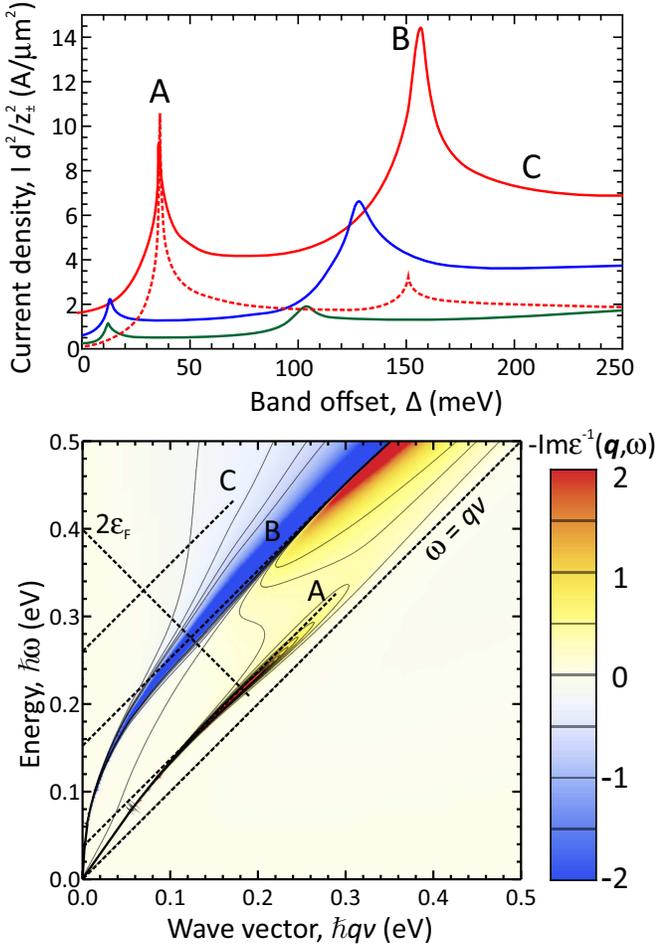}}
    \caption{ (a) Calculated inelastic tunnel current (normalized by $(z_{\pm}/d)^2$) vs band offset $\Delta$ at fixed Fermi energies in graphene layers (red solid $\varepsilon_{F,t}=0.6$ eV, $\varepsilon_{F,b}=-0.2$ eV; blue solid $\varepsilon_{F,t}=0.5$ eV, $\varepsilon_{F,b}=-0.1$ eV; green solid $\varepsilon_{F,t}=0.4$ eV, $\varepsilon_{F,b}=0.1$ eV). Red dashed curve represents the plasmon-assisted current calculated via Eq. (\ref{Plasmon-assisted}) for $\varepsilon_{F,t}=0.6$ eV, $\varepsilon_{F,b}=-0.2$ eV.  Interlayer distance $d=38$ A, $\kappa=5$, temperature $T=300$ K. Peaks A and B correspond to plasmaronic resonances due to acoustic and optical modes, respectively.
   (b) Loss function $-{\rm Im}\left[ \varepsilon^{-1}({\bf q},\omega)\right]$ of the double layer structure for the same parameter values as for red curve in Fig. (a). Resonant peaks in current correspond to the tangent of the interlayer excitations' dispersion $\omega = \Delta + q v$ (dashed line) and dispersion of surface plasmons (bright peaks in the spectral function). The dot-dashed line is the boundary of interband absorption $\omega = 2\min\{\varepsilon_{F,t},\varepsilon_{F,b}\} - qv$}
    \label{fig:IVcurve}
\end{figure}
It is possible to tune the structure parameters to achieve low damping of plasmons with $q=q^*$ by blocking the interband transitions, therefore making the resonant contribution very large. The dependence of full inelastic current on band offset $\Delta$ at fixed carrier densities in the layers is shown in Fig. \ref{fig:IVcurve}. For the sake of generality, we normalize the current by $z^2_{\pm}/d^2$, therefore getting rid of material-dependent tunneling exponent $z_{\pm} \propto e^{-\varkappa d}$, where $\varkappa$ is the decay length of electron wave function. The two peaks in Fig.~\ref{fig:IVcurve} correspond to the plasmaronic resonances on acoustic and optical modes. It is worth noting that the resonance with optical mode is possible only for different conductivities of the two layers. Otherwise, the average mode field between layers is zero and the tunneling matrix elements turn to zero as well.

In the perfectly aligned layers, the inelastic tunneling current is readily seen (and even surpasses) the elastic current, as shown in Fig.~\ref{fig:elastic}. The elastic current scales as $I_{\max}\gamma/[\gamma^2 + \Delta^2]$, where $\gamma$ is the collisional broadening of the resonance. The mentioned effect is, however, highly sensitive to the twist between graphene layers~\cite{Twist-controlled}. With the neglect of emerging weak tunneling between dissimilar sublattices, the general expression for inelastic current (\ref{Many-particle}) still holds, but the interlayer polarizability is now angle-dependent. Denoting the wave vectors connecting the $K$-points in the neighboring layers as $\Delta {\bf q}_i$ ($i = 1...3$), we can write the polarizability in the presence of twist $\Pi^{(T)}_{\pm}({\bf q},\omega)$ as
\begin{equation}
   \Pi^{(T)}_{\pm}({\bf q},\omega) = \frac{1}{3} \sum_{i=1}^{3}{\Pi_{\pm}({\bf q} + \Delta {\bf q}_i,\omega)}.
\end{equation}
When the twist wave vector is small compared to the plasmon wave vector at the resonance, $\Delta q \ll q^*$, the twist-limited contribution to the tunnel current can be estimated as
\begin{equation}
\label{Twist_limit}
    I^{\rm pl,em}_{t\rightarrow b} (\Delta^*) \approx 
          \dfrac{I_0}{2\pi}\sqrt{\dfrac{q_0^2\left.\frac{\partial^2\omega}{\partial q^2}\right|_{q_0}}{v\Delta q}}  \ln\left|\dfrac{q_c^2 \partial^2\omega_q/\partial q_c^2 }{8v\Delta q}\right|.
\end{equation}

\subsection{Plasmon emission and junction electroluminescence}
\begin{figure}
    \centering
    \includegraphics[width=1.0\linewidth]{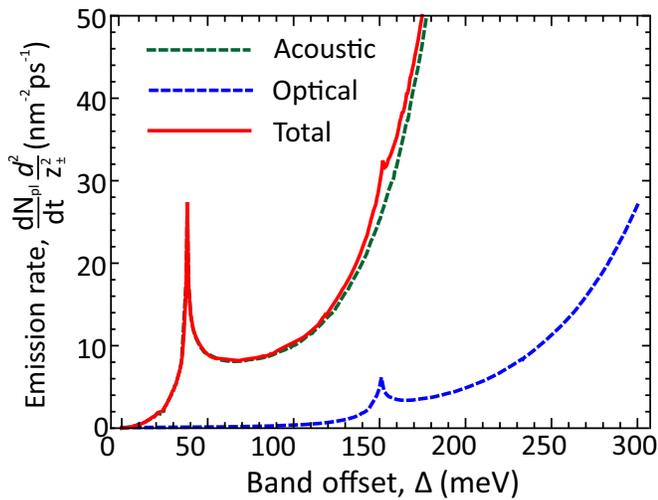}
    \caption{Dependence of plasmon generation rate on offset between Dirac points in graphene layers for the parameters corresponding to red curve in Fig.\ref{fig:IVcurve}a }
    \label{fig:Emission_rate}
\end{figure}

The emission of surface plasmons upon resonant tunneling can be detected not only implicitly by analyzing the features of inelastic current. Recent advances in near-field optical microscopy and electromagnetic sensing at the nanoscale~\cite{Koppens_nano_imaging_hBN,Koppens_thermoelectric} allow a direct measurement of plasmon emission rates. This emission rate, $dN_{pl}/dt$, is obtained by a simple rearrangement of terms in the expression for plasmon-assisted tunnel current
\begin{equation}
        \frac{dN_{pl}}{dt} = \frac{1}{e} \left[ I^{\rm pl,em}_{t\rightarrow b} + I^{\rm pl,em}_{b\rightarrow t} - I^{\rm pl,abs}_{t\rightarrow b} - I^{\rm pl,abs}_{b\rightarrow t} \right].
\end{equation}
Naturally, the bias dependence of integrated plasmon emission rate inherits all resonant features of plasmon-assisted current. This is shown in Fig.~\ref{fig:Emission_rate}, where the characteristic peaks are the plasmaronic resonances discussed above. At large band offsets, the emission will be overwhelmed by absorption, and the double layer structure can operate as a resonant plasmonic photodetector~\cite{Novotny_On-chip_detection,Ryzhii_resonant_detection}. By appropriate choice of doping, it is possible to achieve the plasmaronic resonance both in emission and absorption rates.

It is notable that the energy spectrum of plasmon emission, $dN_{pl,\omega}/dt$, possesses specific resonances at any bias. These resonances reflect the singular nature of interlayer polarizability $\Pi_{\pm}({\bf q,\omega})$ and occur at frequencies satisfying $\omega = \Delta \pm q^p (\omega)$, where $q^p (\omega)$ is the inverse of the dispersion law for the $p$-th mode~\cite{Plasmons_Coupled}. The singularities in the frequency spectrum are integrable except for the case of merging singularities at the plasmaronic resonance.

The spectra of emitted surface plasmons can be studied via analysis of tunnel junction electroluminescence. Such electroluminescence is commonly a two-step process including the excitation of surface plasmon upon inelastic tunneling and its subsequent radiative decay into free-space modes. The direct emission of photon is less probable due to the large spatial extent of photonic modes and small photonic density of states as well. The radiative decay of plasmon with characteristic rate $\gamma_{rad}$ generally competes with damping due to the free carrier scattering and interband absorption having the rate $\gamma_{sc}$. When both decay rates are smaller than plasmon frequency, the plasmon-to-photon conversion rate can be presented as
\begin{equation}
    \frac{dN_{ph}}{dt} = \frac{\gamma_{rad}}{\gamma_{rad}+\gamma_{sc}} \frac{dN_{pl}}{dt}.
\end{equation}

The main obstacle to the observation of plasmonic electroluminescence is the smallness of plasmon radiative damping. Eventually, this is due to the smallness of dipole moment of tunnel junction, and smallness of plasmon velocity $s$ compared to the speed of light. For the tunnel junction of width $W$ and length $L$ tuned to the fundamental mode of surface plasmon $q L = \pi$, the radiative decay rate can be estimated in the dipole approximation to be
\begin{equation}
    \gamma_{rad} \approx \frac{8 L d W \omega^4}{3\pi^3 c^3}\left[2 + \frac{3}{K} + \frac{1}{K^2}\right]^{-1},
\end{equation}
where $K = 4\alpha_c k_F d$, $k_F$ is the Fermi wave vector and $\alpha_c = e^2/\hbar \kappa v$ is the coupling constant. For the acoustic mode of energy $\hbar\omega = 200$ meV (corresponding to the resonance A in Fig. 2), one can estimate $\gamma_{rad}\approx 2 \times 10^8$ s$^{-1}$. This is quite a low value compared to the absorption rate due to electron scattering $\gamma_{sc}\approx 10^{12}$ s$^{-1}$~\cite{Koppens_nano_imaging_hBN}. The resulting efficiency of plasmon-to-photon conversion is about $\gamma_{rad}/\gamma_{sc} \approx 2\times 10^{-4}$. 

This efficiency can be increased dramatically by coupling a resonant-tunneling structure with a nanoscale antenna~\cite{Parzefall2017}. An estimate for the radiative decay rate in this case can be obtained by solving the plasmon dispersion equation in a bounded double-layer structure loaded by an antenna impedance $Z_{rad}$. The maximum decay rate is achieved when the load impedance matches the impedance of a double-layer structure at the resonant frequency,
\begin{equation}
    Z_{rad} \sigma''(\omega_{\bf q}) W/L \sim 1,
\end{equation}
the definite value of numerical factor depends on the shape and biasing of the double layer and is evaluated in Appendix B. Under optimal loading condition, the radiative decay rate becomes the same order as the eigenfrequency
\begin{equation}
    \gamma_{rad,\max} \sim \omega'_{\bf q},
\end{equation}
a numerical estimate for the structure with equal layer doping yields $\gamma_{rad,\max} =0.04 \omega'_{\bf q}$ for the lowest mode. As the scattering rate is well below the resonant frequency, the plasmon-to-photon conversion efficiency can be made close to unity. 

\subsection{Fine structure of the low-temperature $I(V)$-curves}
\begin{figure}
    \centering
    \includegraphics[width=1.0\linewidth]{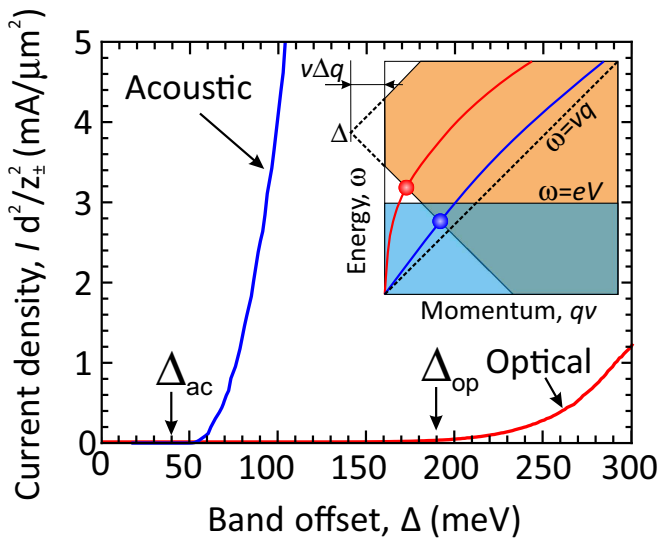}
    \caption{Calculated dependence of plasmon-assisted tunnel current at fixed carrier densities in the layers ($\varepsilon_{F+}=225$ meV, $\varepsilon_{F-}=250$ meV) vs band offset $\Delta$ at $T=0$ and $\Delta q=0$. The offsets $\Delta_{ac}$ and $\Delta_{op}$ correspond to the switch-on of plasmon-aided tunneling with emission of acoustic and optical plasmons, respectively. An inset shows the diagram for geometrical determination of threshold voltage: the intraband tunneling with acoustic (optical) plasmon emission becomes possible when blue (red) circle appears below the Pauli blocking line $\omega = eV$. For the conditions shown in inset, the emission of acoustic plasmons is possible while emission of optical is not}
    \label{fig:Cusps}
\end{figure}

At low temperatures, the tunneling with plasmon absorption if frozen out, while emission aided tunneling is possible only for $\omega_q < eV$ by the virtue of Pauli blocking. The combination of Pauli restriction and energy-momentum conservation results in suppression of inelastic current for certain range of voltages $V < V_{th}$ and band offset $\Delta < \Delta_{th}$, while above the threshold value the inelastic current switches on in a threshold-like manner. The threshold structure of the plasmon-aided current repeats that of an integral
\begin{equation}
\label{Fine_struct}
    J(V,\Delta) = \sum_{{\bf p}{\bf q}ss'}{\theta(eV - \omega_{\bf q})\delta(\epsilon^{s+}_{\bf p+}-\epsilon^{s'-}_{\bf p+} - \omega_{\bf q})},
\end{equation}
where ${\bf p}_{\pm} = {\bf p}\pm ({\bf q}+\Delta {\bf q})/2$ are the momenta of initial and final electrons, the Heaviside theta function is responsible for Pauli blocking, and the delta-function for the energy conservation. The intraband tunneling transitions are  stronger, and therefore we consider the terms with $s=s'$ only. Evaluation of integral (\ref{Fine_struct}) leads us to the following threshold condition
\begin{gather}
\label{Threshold}
eV_{th} = \omega_{q},\\
\omega_{q} = \Delta_{th}-(q+\Delta q)v.
\end{gather}
The latter admits a simple geometrical interpretation shown in the inset of Fig.~(\ref{fig:Cusps}). The minimal frequency of plasmon in the domain of intraband tunneling $|\omega - \Delta| < (q+\Delta q)v$ (orange filled region) should lie below the line of Pauli blocking $\omega = e V$ (blue filled region). From this analysis we also see that finite interlayer twist $\Delta q$ reduces the threshold of plasmon emission upon tunneling.

For acoustic plasmon in graphene double layer with linear dispersion $\omega = s q$, the threshold condition (\ref{Threshold}) can be solved analytically to yield
\begin{equation}
\label{Ac-threshold}
 e V_{th} = \frac{\Delta - v\Delta q}{1+v/s}.
\end{equation}
If the band offset $\Delta$ is fixed, the threshold voltage (\ref{Ac-threshold}) weakly depends on carrier density because the plasmon velocity $s$ tends to the Fermi velocity at small interlayer distance $d$~\cite{Ryzhii-plasmons,Polini-soundarons,Plasmons_Coupled}, and the density dependence of $s$ is weak. This contrasts to the case of plasmon-assisted tunneling in bulk metal-insulator-metal junctions~\cite{Lambe_PRL} where the threshold voltage equals the plasmon energy, the latter scaling as square root of density.

If the carrier densities in graphene layers are fixed while band offset is swept, the threshold condition can be presented in an alternative form
\begin{equation}
 \Delta_{th} = \left( 1+\frac{s}{v} \right) (\varepsilon_{F-}-\varepsilon_{F+}) - s \Delta q.
\end{equation}

The step-like switch-on of tunnel current upon increase in band offset $\Delta$ is shown in Fig.~(\ref{fig:Cusps}) for fixed carrier densities and zero temperature. These cusps in the $I(V)$-curves become broadened very quickly with the temperature increase, as the broadening is governed both by smearing of Fermi distributions and interband damping of graphene plasmons.


\section{Discussion and conclusions}
We have theoretically identified several manifestations of plasmon-assisted tunneling in graphene-insulator-graphene junctions, the most striking of them being the plasmaronic resonance in tunnel current. The origin of this resonance is the enhanced interaction between plasmons and interlayer single-particle excitations due to the group velocity matching. The relation between discussed resonance and formation of plasmarons in a single graphene layer is elucidated as follows. Plasmaron is formed off the mass shell $\epsilon_{p} = p v_0$ at some energy separation $\delta \epsilon$ equal to the energy of plasmon quantum $\omega_q$. Contrary to 3D systems, the plasmons in two-dimensions have soft spectrum with energy tending to zero at long wavelength. A natural question arises: which plasmon wave vector $q^*$ provides the strongest interaction with electrons? The answer is that such a plasmon should have group velocity equal to the carrier velocity\cite{Bostwick}. 
In the case of interlayer tunneling, we are dealing essentially with on-shell electrons, however, the energy of interlayer single-particle excitations is tuned by interlayer bias $\Delta$. At some bias $\Delta^*$, the energies, momenta and group velocities of interlayer excitations and plasmons coincide. This bias corresponds to the resonantly large generation of surface plasmons by interlayer tunneling.

We note that the mentioned resonance is closely related to the presence of square-root singularities in the interlayer polarizability of double layer. Such singularities are inherent to the linear carrier dispersion in graphene~\cite{Das_Sarma_Plasmons} and are absent in systems with parabolic bands~\cite{Stern_2degPolarizability}. Similar singularities in the polarizability of a single layer manifest themselves in a number of interesting phenomena, including ultrafast collinear scattering of photoexcited carriers~\cite{Malic_SlowNoncollinear} and the absence of Landau damping for plasmons~\cite{Ryzhii-plasmons}. These singularities can be smeared by electron-electron corrections to carrier dispersion and/or by vertex corrections~\cite{New_plasmon_mode}. Though our original derivation of inelastic current was based on the scattering of noninteracting particles, the interaction effects can be conveniently included in the transformed equation~(\ref{Many-particle}) by replacing the bare polarizabilities $\Pi_{ll'}$ with interacting ones. Here we just note that logarithmic interaction corrections to linear bands are most relevant for quasi-neutral graphene, while for doped samples the e-e interactions just enhance the band velocity under the Fermi surface~\cite{Hwang_DensityDependentExchange}. In most realistic situations, the broadening of predicted plasmaronic resonance would be governed by finite plasmon lifetime, Eq.~(\ref{Scattering_limit}), and interlayer twist, Eq.~(\ref{Twist_limit}).

In the present calculation, we assumed the interlayer tunneling to be weak, so that the dielectric function of the double layer was not renormalized by tunneling. Such renormalization can be done~\cite{Plasmons_Coupled}, and it would enhance the plasmon-assisted current. The reason for enhancement is the partial plasmon loss compensation by stimulated plasmon emission upon tunneling. At some critical strength of tunneling, corresponding to the complete undamping of plasmon modes, the current (\ref{Many-particle}) would diverge. This effect is analogous to the enhancement of scattering rates in photoexcited semiconductors due to plasmon undamping~\cite{Scott_Undamping}. The divergence would signalize on the onset of surface plasmon lasing, at this point one has to solve the coupled kinetic equations for electrons and plasmons for evaluation of tunnel current.

The present theory demonstrates the prospect of graphene heterostructures for resonant and voltage tunable light emission in the far infrared. Compared to the light sources based on tunneling injection and recombination of excitons~\cite{Exciton_recombination}, the proposed structures can be scaled down to the nanometre dimensions due to high confinement of plasmons, and integrated in photonic and plasmonic waveguides. The proposed process of plasmon and photon generation is also different from interband recombination of electrons injected upon resonant tunneling considered in~\cite{Khorasani}. The difference between these processes is the same as the difference of quantum-cascade lasing with vertical and diagonal radiative transitions. Importantly, the emission spectrum for diagonal transitions is resonant due to the singularities in the {\it joint} density of states between layers, while the emission spectrum for interband recombination of injected carriers repeats the smooth spectrum of emission in a single layer.

In conclusion, we have developed a theoretical formalism for the calculation of tunneling current accompanied by carrier-carrier scattering in graphene-insulator-graphene heterostructures. Our calculation shows that the main contribution of inelastic scattering-assisted current comes from emission of surface plasmons. The plasmon-assisted current can be resonantly enhanced if the energy, momentum and group velocity of interlayer excitations and plasmons coincide. This effect, which we call plasmaronic resonance, can also manifest itself in enhanced plasmon emission and electroluminescence of graphene-based junctions.

\begin{acknowledgments}
This work was supported by the grants \# 16-37-60110/16 and \# 16-29-03402/16 of the Russian Foundation for Basic Research and by the Grant of the President of the Russian Federation. The authors are grateful to G. Alymov for valuable discussions.
\end{acknowledgments}

\appendix

\section{Second-quantization approach to plasmon-assisted tunneling}
\label{AppA}
The plasmon-assisted tunneling current can be calculated as a current induced by random fluctuations of electric potential
\begin{equation}
    \varphi\left(\bm{r}\right)=\frac{1}{2}\sum_{{\bf q},p}\varphi^{p}_q(z)e^{i{\bf q r} - i\omega t} + c.c.
\end{equation}
Using the Fermi golden rule, we present this current as
\begin{equation}
    I^{pl, em}_{t\rightarrow b} = {2\pi e} \sum_{{\bf p}{\bf q}ss'}{f^{+s}_{\bf p} (1 - f^{-s'}_{\bf p - q}) |e\varphi^{em}_{tb}|^2 \delta(\varepsilon^{+s}_{\bf p} - \varepsilon^{-s'}_{\bf p - q} - \omega_q) }
\end{equation}
for emission contribution, and 
\begin{equation}
    I^{pl, abs}_{t\rightarrow b} = {2\pi e} \sum_{{\bf p}{\bf q}ss'}{f^{+s}_{\bf p} (1 - f^{-s'}_{\bf p + q}) |e\varphi^{abs}_{tb}|^2 \delta(\varepsilon^{+s}_{\bf p} - \varepsilon^{-s'}_{\bf p + q} + \omega_q) }
\end{equation}
for absorptive contribution. For classical field, the matrix elements are the same for emission and absorption. For quantized field, one has
\begin{gather}
    |e\varphi^{em}_{tb}|^2 = [N_{\omega^{p}_q} + 1]|e\varphi_{tb}|^2,\\
    |e\varphi^{abs}_{tb}|^2 = N_{\omega^{p}_q} |e\varphi_{tb}|^2.
\end{gather}
The tunneling matrix element is calculated as
\begin{equation}
    e\varphi_{tb} = \int_{-\infty}^{+\infty}{dz \psi^*_{t}(z) e\varphi_{\bf q}(z) \psi_b(z)},
\end{equation}
where $\psi^*_{l}(z)$ is the $z$-component of the wave function belonging to the $l$-th layer, and  $\varphi^{p}_q(z)$ is the normalized electric potential in the plasmon mode with wave vector ${\bf q}$ and polarization $p$. The field magnitude can be parametrized by a single quantity $\varphi_0$, while the spatial dependence of electric potential is given by:
\begin{gather}\label{plasmon_profile}
    \varphi_{{\bf q}}(z)=\varphi_0 S_q(z),\\
    S_q(z) =
    \begin{cases}
    S_+ e^{-q(z-d/2)}, \quad z>d/2 \\
  \frac{ S_+ \sinh[q(z+\frac{d}{2})] - S_- \sinh[q(z - \frac{d}{2})] }{\sinh qd},\quad |z|<d/2,\\
    S_- e^{q(z+d/2)}, \quad z<-d/2,
    \end{cases}
\end{gather}
with \mbox{$S_+=e^{-qd}$}, \mbox{$S_-=1+i2\pi q\sigma''_{+}({\bf q},\omega_q^{p})\left(1 - e^{-2qd}\right)/\kappa\omega_q$}, \newline$\sigma_{+/-}({\bf q},\omega_q^{p})=i\omega_q e^2\Pi_{++/--}\left({\bf q},\omega_q\right)/q^2$ is the conductivity of top/bottom graphene layer. To determine the amplitude of plasmon field $\varphi_0$ we equate the classical field energy in a dispersive medium to the quantized energy $\hbar\omega^p_{\bf q}$. The classical energy of electromagnetic field is
\begin{multline}
\label{Brillouin_formula}
    W =\int d^3r\frac{\kappa {\bf E}{\bf E}^*}{16\pi}- \frac{A}{4}\sum_{l=\pm}\left.\frac{\partial\sigma''_{l}}{\partial\omega}\right|_{\omega_q} \left.{\bf E}_{||}{\bf E}_{||}^*\right|_{z=ld/2}=\\
     =\dfrac{A\varphi_0^2q\omega_qS_-}{8\pi}\left.\dfrac{\partial\varepsilon}{\partial\omega}\right|_{\omega_q},
     \end{multline}
where \mbox{${\bf E}=\left({\bf E}_{||},E_z\right)=-\left(i{\bf q},\partial_z\right)\varphi_q(z)$} is the electric field in plasmon mode and $A$ is the sample area. This leads us to the final formula for $\varphi_0$:
\begin{equation}
    \left( \dfrac{e \varphi_0}{2}\right)^2 =\dfrac{\hbar V_0({\bf q})} {A S_- \left.\partial\varepsilon / \partial\omega \right|_{\omega_q}},
\end{equation}
here we have introduced the permittivity of the double layer structure:
\begin{align}\label{dielectric_function}
\varepsilon(\bm{q},\omega) = \left[1-V_0(\bm{q})\Pi_{++}(\bm{q},\omega)\right]
\left[1-V_0(\bm{q})\Pi_{--}(\bm{q},\omega)\right] - \nonumber\\
-e^{-2 qd}V_0(\bm{q})^2\Pi_{++}(\bm{q},\omega)\Pi_{--}(\bm{q},\omega).    
\end{align}

\section{Plasmon poles in scattering-assisted tunneling}
\label{AppB}
The current accompanied by emission of plasmons can be also derived by extracting the contribution to the integral (\ref{Many-particle}) due to the poles of screening function $\varepsilon^{-1}({\bf q},\omega)$. Assuming the dissipation of electromagnetic energy to be small, one can determine the plasmon frequency $\omega_q^p$ from:
\begin{align}\label{plasmon_disp}
\varepsilon'(\bm{q},\omega_q^p) = 0.    
\end{align}
If the frequency $\omega$ and momentum ${\bf q}$ satisfy the disperion relation (\ref{plasmon_disp}), the transition amplitudes $V_{++,-+}$ with $V_{+-,--}$ are related as follows:
\begin{equation}
\label{amplitude_relat}
    V_{++,-+}=V_{+-,--}\frac{V_0(\bm{q})\Pi'_{--}(\bm{q},\omega_q)e^{-qd}}{1-V_0(\bm{q})\Pi'_{++}(\bm{q},\omega_q)}.
\end{equation}
With the help of Eqs. (\ref{plasmon_disp}),(\ref{amplitude_relat}) we can write down the tunneling current in the following way (keeping in mind $\omega\approx\omega_q$):
\begin{align}\label{current_transformation}
I_{t \rightarrow b}=\frac{2e}{\pi} \int\limits_{-\infty }^{+\infty }{d\omega }\sum\limits_{{\bf q},l} \Pi''_{+-}( q, \omega -\Delta)N_{\omega -eV} \left( N_\omega + 1 \right)\times \nonumber\\
\times\left[\left| V_{++,-+} \right|^2\Pi''_{--}( \bm{q}, \omega) + \left| V_{+-,--} \right|^2\Pi''_{++}( \bm{q}, \omega)\right]  = \nonumber\\
\frac{2e}{\pi} \int\limits_{-\infty }^{+\infty }{d\omega }\sum\limits_{{\bf q},l} \Pi''_{+-}( q, \omega -\Delta)N_{\omega -eV} \left( N_\omega + 1 \right)\times \nonumber\\
\times\left| V_{+-,--} \right|^2\frac{V_0(\bm{q})\left|\Pi'_{--}( \bm{q}, \omega)\right|}{\left|1-V_0(\bm{q})\Pi'_{++}( \bm{q}, \omega)\right|}\left[-\varepsilon''(\bm{q},\omega)\right] 
\end{align}
In the limit $|\varepsilon''/\varepsilon'|\ll 1$ one can make a substitution
\begin{equation}
    \frac{\varepsilon''(\bm{q},\omega)}{\left|\varepsilon(\bm{q},\omega)\right|^2} \approx
    2\pi \sum_{p=\pm 1}
    \frac
    {\delta(\omega-\omega_q^{p}) + \delta(\omega+\omega_q^{p}) } 
    {\left|\left.\partial\varepsilon'/\partial\omega \right|_{\omega_q^p}\right|}
\end{equation}
in the Eq. (\ref{current_transformation}) after the last equality and obtain the resulting expression for the tunneling current:
\begin{align}\label{plasmon_assisted_appendix}
I_{t \rightarrow b} = 4e \sum\limits_{{\bf q},p=\pm 1}\left| S^{p}_{+-}(q) \right|^2\frac{V_0^2(\bm{q})\left|\Pi'_{--}( \bm{q}, \omega_q^{p})\right|}{\left|1-V_0(\bm{q})\Pi'_{++}( \bm{q}, \omega_q^{p})\right|}\times \nonumber \\
\left[\Pi''_{+-}( q, \omega_q^{p} -\Delta)N_{\omega_q^{p} - eV} \left( N_{\omega_q^{p}} + 1 \right) + \right. \nonumber \\
\left.+\Pi''_{+-}( q, \omega_q^{p} + \Delta)N_{\omega_q^{p}} \left( N_{\omega_q^{p} + eV} + 1 \right) \right], 
\end{align}
where 
\begin{align}
S_{+-}^{p}(q)=V_0(\bm{q})\Pi'_{++}( \bm{q}, \omega_q^{(p)})e^{-qd}e^{-q|z-d/2|} + \nonumber \\
+\left( 1 - V_0(\bm{q})\Pi'_{++}( \bm{q}, \omega_q^{(p)})\right)e^{-q|z+d/2|} 
\end{align}
is a dimensionless potential profile which equals the plasmon potential $S_q(z)$ (\ref{plasmon_profile}).

\section{Tunneling matrix elements}
Electron states in coupled graphene layers with small interlayer twist can be described using the Hamiltonian
\begin{equation}
\label{Hamiltonian}
\hat{H}_0=\left( \begin{matrix}
   {{{\hat{H}}}_{G+}} & {\hat{\mathcal{T}}}  \\
   {{{\hat{\mathcal{T}}}}^{*}} & {{{\hat{H}}}_{G-}}  \\
\end{matrix} \right),
\end{equation}
where the blocks ${{\hat{H}}}_{G\pm}$ describe isolated graphene layers, and $\hat{\mathcal{T}}$ describes tunnel hopping:
\begin{equation}
\label{Tunneling}
{\hat{\mathcal{T}}} = \frac{\Omega}{3} \sum_{j=0,1,2} {e^{-i \Delta {\bf q}_j {\bf r}}\left( \begin{matrix}
   1 & e^{-i \frac{2\pi j}{3}}  \\
   e^{i \frac{2\pi j}{3}} & 1  \\
\end{matrix} \right)}.
\end{equation}
Here $\Omega$ is the tunneling overlap integral and $\Delta {\bf q}_j$ are the wave vectors connecting the respective edges of hexagonal Brillouin zones in the layers. In the absence of interlayer twist ($\delta {\bf q}= 0$), the tunneling matrix is diagonal, ${\hat{\mathcal{T}}} =\Omega {\hat I} $, where ${\hat I}$ is the identity matrix. In this case, the band and layer degrees of freedom are decoupled.

We proceed now to the evaluation of matrix element $\Omega$. The physical meaning of $\Omega$ is half the energy splitting between electron states in coupled graphene layers, as can be seen from diagonalization of Hamiltonian (\ref{Hamiltonian}). On the other hand, this splitting can be estimated from a continuum model, where each graphene layer is represented by a delta-well~\cite{Plasmons_Coupled}. The delta-well potential is
\begin{equation}
    U(z) = 2\sqrt{\frac{\hbar^2 U_0}{2m^*}}[\delta(z-d/2) + \delta( z + d/2) ],
\end{equation}
where $U_0$ is the work function from graphene to the barrier material, and $m^*$ is the effective mass in the barrier. For boron nitride, $U_0 \approx 1.5$ eV and $m^* \approx 0.5 m_0$. The eigen functions in this potential are symmetric and anti-symmetric ones. The energy difference between these states is 
\begin{equation}
   E_+ - E_- = 2\Omega = 4 U_{0} e^{-\varkappa d},
\end{equation}
where $\varkappa = \sqrt{2m^* U_0}/\hbar$ is the decay constant of electron wave function.



The plasmon-assisted current is proportional to the matrix element of electric potential energy $e\varphi_{\pm}$. Its evaluation generally requires the knowledge of electron wave function inside the barrier layer. This evaluation can be, however, simplified in the dipole approximation. We write the potential distribution in the plasmon mode as $\varphi_q(z) = {\bar \varphi} + (\varphi_+ - \varphi_-) z/d$, thus the potential matrix element becomes:
\begin{equation}
    e\varphi_{\pm} \approx (\varphi_+ - \varphi_-) \frac{z_{\pm}}{d}.
\end{equation}

It appears that the coordinate matrix element $z_\pm$ and the tunnel splitting $\Omega$ are bound by a simple relation. We consider two methods for calculation of current between states $\ket{+}$ and $\ket{-}$. From one hand, this can be expressed through velocity operator in the transverse direction:
\begin{equation}
    j_{\pm} = \frac{(v_z)_{\pm}}{d} = \frac{z_{\pm}}{\hbar d} (\epsilon_+ - \epsilon_-) = \frac{z_{\pm}}{\hbar d}  \sqrt{\Delta^2 +4\Omega^2}.
\end{equation}
From the other hand, it can be found by evaluating the derivative of particle number in the state $\ket{+}$
\begin{equation}
j_{\pm} = \frac{d N_+}{d t} = -\frac{i}{\hbar}[\hat N_+, \hat H_0] = \frac{\Omega}{\hbar}.   
\end{equation}
Comparing these two expressions, we find
\begin{equation}
z_{\pm} = d \frac{\Omega}{\sqrt{\Delta^2 + 4\Omega^2}}.
\end{equation}


\begin{figure}
    \centering
    \includegraphics[width=8cm,height=5cm]{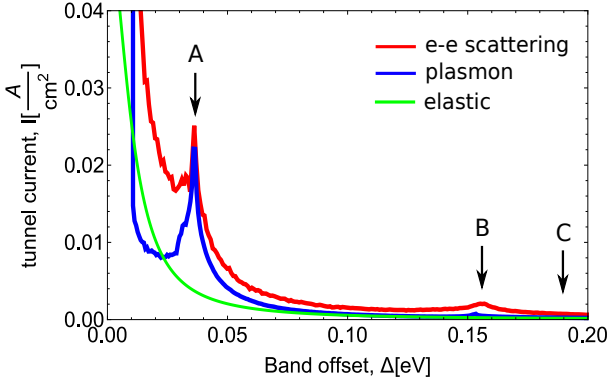}
    \caption{Dependence of absolute values of tunnel currents on band offset at $\varepsilon_{F,+}=0.6$ eV, $\varepsilon_{F,-}=-0.2$ eV, $d=38$ A. We use broadening $\gamma=10$ meV for elastic current.}
    \label{fig:elastic}
\end{figure}

\section{Comparison of elastic and inelastic currents}
The elastic current can be evaluated by considering the tunneling matrix elements in Hamiltonian (\ref{Hamiltonian}) as small perturbations. This leads to the following formula~\cite{Brey_PRA}
\begin{multline}
\label{Elastic}
    I^{el}=\frac{g e}{\hbar}\sum_{\bm{k}ss'}\int_{-\infty}^{+\infty}\frac{dE}{2\pi} \left|\Omega^{+s,-s'}_{{\bf k},{\bf k}+\Delta q} \right|^2 \times \\
    A_{+s}({\bf k},E) A_{-s'}({\bf k}+\Delta {\bf q},E)\left[f(E) - f(E - eV) \right],
\end{multline}
here $A_{l s}({\bf k},E) = -2 {\rm Im} G^R_{ls}({\bf k}, E)$ is the spectral function in the $l$-th layer and $s$-th band, and $G^R_{ls}({\bf k}, E)$ is the retarded Green's function in graphene in the band representation:
\begin{equation}
    G^R_{ls}({\bf k}, E) = [E - \epsilon^{ls}_{\bf k} - \Sigma^{ls}({\bf k},E) + i \delta]^{-1},
\end{equation}
$\Sigma^{ls}({\bf k},E)$ is the electron self energy. In the simplest approximation, the self energy can be treated as a constant $\Sigma^{ls}({\bf k},E) \approx \gamma$. In this case, one can approximate
\begin{multline}
    I^{el} \approx \frac{2\pi g e}{\hbar}\sum_{\bm{k}ss'} \left|\Omega^{+s,-s'}_{{\bf k},{\bf k}+\Delta q} \right|^2 \delta_\gamma(\epsilon^{+s}_{\bf k} - \epsilon^{-s'}_{\bf k + \Delta q}) \left[f^{+s}_{\bf k} - f^{-s'}_{\bf k+\Delta q} \right],
\end{multline}
where $\delta_\gamma(x) = (\gamma/\pi)/(\gamma^2 + x^2)$ is the ''broadened'' delta-function. If, in addition, the layers are aligned ($\Delta {\bf q} \rightarrow 0$), the integration is performed trivially yielding the electron ($n$) and hole ($p$) densities in the layers
\begin{equation}
    I^{el} \approx \frac{2\pi e}{\hbar}|\Omega|^2 \delta_\gamma(\Delta) \left[ (n_+ - p_+) - (n_- - p_-) \right].
\end{equation}

\section{Radiative decay of plasmon modes}
The present section is aimed at the estimate of plasmon-to-photon conversion efficiency. The plasmon emitted upon tunneling can decay either radiatively (with the rate $\gamma_{rad}$) or it can be re-absorbed due to the Drude or intraband absorption in a single layer (the corresponding rate is $\gamma_{abs}$). Considering these competing channels of plasmon decay, we can estimate the plasmon-to-photon conversion probability as $\gamma_{rad}/(\gamma_{rad} + \gamma_{abs})$.

The radiative decay rate of plasmon $\gamma_{\rm rad}$ can be estimated as
\begin{equation}
    \gamma_{\rm rad} = \frac{P_{\rm rad}}{W},
\end{equation}
where $P_{\rm rad} = (4\omega^4/3c^3) |{\bf d}_\omega|^2$ is the power of dipole radiation. We consider a double graphene layer sample of length $L$ and width $W$, so that $L$ corresponds to the fundamental plasmon mode $q L = \pi$. As example, we consider an acoustic plasmon mode with linear dispersion $\omega = s q$. Evaluating the average dipole moment, we find the radiated power
\begin{equation}
\label{Power}
    P = \frac{1}{3c^3}\left(\frac{\omega^2 \kappa L W}{\pi^2}\right)^2 \varphi^2_0.
\end{equation}
The mode energy is
\begin{equation}
\label{Energy}
    W = \frac{\kappa}{8\pi}q^2\varphi^2_0\left[1+\coth\frac{qd}{2}\right] - \frac{1}{2}\frac{d\sigma''}{d\omega}q^2\varphi^2_0,
\end{equation}
where the first term is due to field and the second one is due to the particle motion. Further estimates will be done in the quasi-classical limit, $qv\ll \varepsilon_F$, $\omega \ll \varepsilon_F$. In this approximation, the in-plane conductivity is essentially intraband:
\begin{equation}
\label{InPlane-cond}
    \sigma'' = g \frac{{e^2}}{\hbar }\frac{\varepsilon_F}{2\pi \hbar}\frac{\omega }{q^2v_0^2}\left[ \frac{\omega }{\sqrt{{\omega^2}- q^2 v_0^2}}-1 \right].
\end{equation}
The solution for acoustic plasmon dispersion law with conductivity (\ref{InPlane-cond}) leads to $\omega = s q$ with the velocity
\begin{equation}
    s = v \frac{1 + K}{\sqrt{1 + 2 K}},
\end{equation}
where $K = 4\alpha_c k_F d$, $\alpha_c = e^2/\kappa\hbar v$ is the coupling constant and $k_F$ is the Fermi wave vector. Evaluating the frequency derivative of conductivity (\ref{InPlane-cond}) at the dispersion curve $\omega = s q$, we arrive at the following expression for mode energy in the long-wavelength limit
\begin{equation}
\label{Energy}
    W = \frac{\kappa}{8\pi}q^2\varphi^2_0 \left[1 + \frac{2}{qd}\left(2+\frac{3}{K}+\frac{1}{K^2}\right)\right].
\end{equation}
Combining Eqs.~(\ref{Power}) and (\ref{Energy}), we find the radiative decay rate
\begin{multline}
\gamma_{rad} = \frac{16}{3\pi^4}\frac{L^2 W \omega^4}{c^3} \left[1 + \frac{2}{qd}\left(2+\frac{3}{K}+\frac{1}{K^2}\right)\right]^{-1}\approx\\
\frac{8}{3\pi^3}\frac{L d W \omega^4}{c^3} \left[ 2+\frac{3}{K}+\frac{1}{K^2}\right]^{-1}.
\end{multline}
We note that the radiative decay rate of plasmons is generally small due to two reasons: (1) smallness of the dipole moment of double layer structure (the factor of $d$ in the numerator) (2) smallness of plasmon velocity compared to the velocity of light (the $c^3$-term in the denominator).

\section{Antenna coupling of plasmons}
The plasmon-to-photon conversion efficiency can be markedly increased if the double layer device is loaded with an antenna. To model the plasmon decay in this situation, we consider the two graphene layers connected via the radiative resistance $Z_{rad}$. We shall solve the dispersion equation for plasmons in this structure and find their decay rate due to radiation. An electric potential is sought for as a superposition of forward and backward optical and acoustic waves
\begin{figure}
    \centering
    \includegraphics[width=0.95\linewidth]{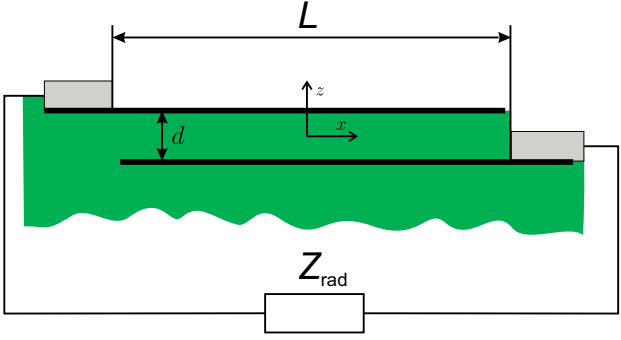}
    \caption{Schematic of double layer graphene structure loaded by an antenna with radiative resistance $Z_{rad}$}
    \label{fig:Structure}
\end{figure}
\begin{equation}
\label{Potential}
\varphi_{\pm} = a e^{i q_+ x} + b e^{-i q_+ x} \pm c e^{i q_- x} \pm d e^{-i q_- x},
\end{equation}
where $q_-$ and $q_+$ are the wave vectors of acoustic and optical modes. The boundary conditions for the schematic in Fig.~\ref{fig:Structure} are
\begin{gather}
\label{BC}
\left. \frac{\partial {\varphi_+}}{\partial x} \right|_{L/2}=\left. \frac{\partial {\varphi_-}}{\partial x} \right|_{-L/2}=0,\\
\left.{\varphi_+} \right|_{-L/2}=-\left. {\varphi_-} \right|_{L/2} = \frac{1}{2} I Z_{rad},
\end{gather}
where $I$ is the current induced in external circuit. Solving Eq. (\ref{Potential}) with boundary conditions (\ref{BC}), we obtain the following dispersion relation
\begin{multline}
\label{General_dispersion}
1 + \cos q_+L \cos q_-L -\frac{1}{2}\left[\frac{q_+}{q_-} + \frac{q_-}{q_+} \right] \sin q_-L \sin q_+ L = \\
\sigma Z_{rad} \frac{W}{L} \left[\frac{q_-}{q_+} \sin q_+ L \sin q_-L - \sin^2\frac{q_-L}{2}\sin^2\frac{q_+L}{2}\right].
\end{multline}
It is possible to estimate the solutions analytically in the limit $q_+/q_- \ll 1$, i.e. the wavelength of optical plasmon much exceeds that of acoustic one. This is generally fulfilled as the acoustic plasmons have linear dispersion while the optical have a square-root one (see also Fig. 2). In this limit, the general dispersion equation (\ref{General_dispersion}) is decoupled into two, neither depending on $q_+$:
\begin{gather}
\cos\frac{q_-L}{2} = 0,\\
\label{Disp_simple}
\frac{q_-L}{2}\tan\frac{q_-L}{2} = \frac{1}{ 1 + 2 \sigma Z_{rad} W / L}.
\end{gather}
Only the solutions of the second equations are affected by the radiative decay. It is convenient to rewrite it introducing the dimensionless frequency $ u = q_- L/2 = \omega L/2 s$, and the dimensionless radiative resistance
\begin{equation}
{\tilde Z} = Z_{rad} \frac{W}{L} \frac{e^2}{\hbar}\frac{\varepsilon_F}{\hbar \omega_{pl}},
\end{equation}
where $\omega_{pl} = \pi s/L$. The dispersion equation becomes
\begin{equation}
 u \tan u = \frac{1}{1 - i {\tilde Z}/u}.
\end{equation}
A general feature of its solutions is that the imaginary part of frequency has an extermum as a function of $\tilde Z$. This is illustrated in Fig.~(\ref{fig:RadDecay}) which shows the decay rate of five lowest plasmon modes vs. radiative resistance $Z_{rad}$ calculated with numerical solution of Eq.~(\ref{Disp_simple}).
\begin{figure}
    \centering
    \includegraphics[width=0.95\linewidth]{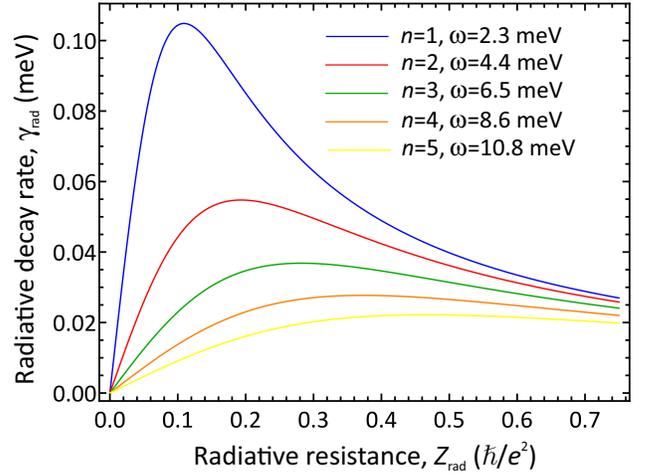}
    \caption{Radiative decay rate of plasmon modes supported by the double graphene layer vs. antenna radiative resistance $Z_{rad}$ (measured in units of $\hbar/e^2$). Channel length $L=2$ $\mu$m, channel width $W=L$, interlayer distance $d=2.5$ nm, Fermi energy $\varepsilon_{F+}=\varepsilon_{F-} = 75$ meV, $T=300$ K}
    \label{fig:RadDecay}
\end{figure}
There exists an optimal value of antenna resistance providing the maximum radiative decay rate. Decoupling the solutions into real and imaginary parts, $u = u' + i u''$, we find that the maximum of decay rate is achieved if
\begin{gather}
{\tilde Z} = u',\\
u'' \approx \frac{\cos^2 u'}{2u'}.
\end{gather}
For the two lowest modes we have obtained $u' = 3.4$, $u'' = -0.14$ and $u'=6.4$, $u''=-0.08$, respectively. We note that the condition of maximum radiative decay ${\tilde Z} = u' \sim 1$ represents the matching of antenna impedance and impedance of graphene layer at the resonant plasmon frequency. For the lowest mode, the maximum decay rate is $\gamma_{rad} \approx 0.04 \omega$. This greatly exceeds the decay rate due to dipole radiation into free space.
\begin{figure}[ht!]
    \centering
    \includegraphics[width=0.95\linewidth]{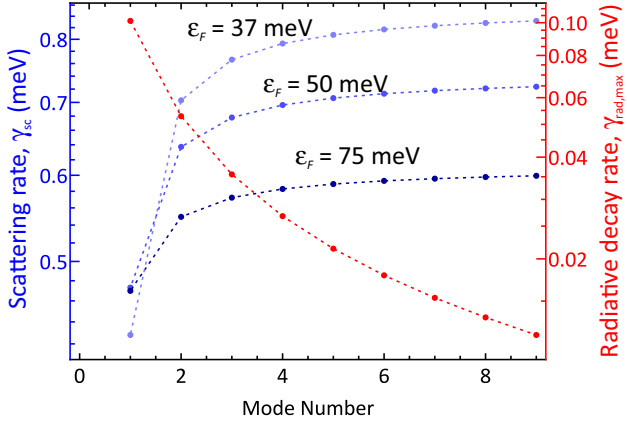}
    \caption{Comparison of radiative ($\gamma_{rad}$) and scattering ($\gamma_{sc}$) decay channels for plasmons supported by the double layer (structure parameters as in Fig.~\ref{fig:RadDecay}) at different Fermi energies. $\gamma_{rad}$ is calculated for antenna resistance $Z_{rad}$ providing the maximum radiative decay. The scattering rate $\gamma_{sc}$ is limited by graphene acoustic phonons and residual charge impurities with density $N_i = 10^{11}$ cm$^{-2}$. Dashed lines are just a guide for eye}
    \label{fig:RadvsScattering}
\end{figure}

\begin{figure}[ht]
    \centering
    \includegraphics[width=0.95\linewidth]{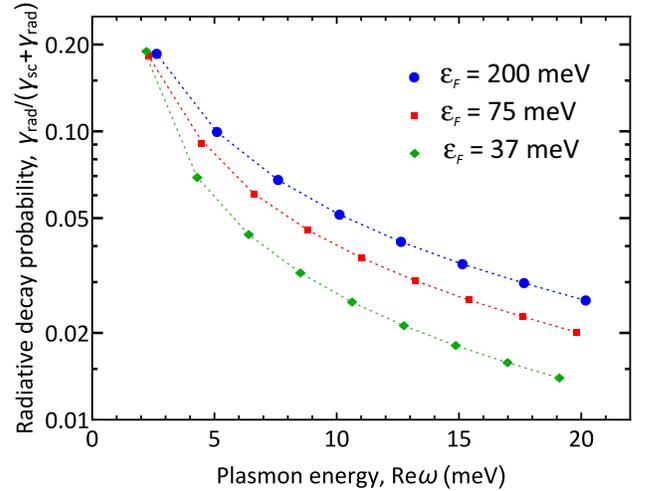}
    \caption{Probability of plasmon decay into free space modes $\gamma_{rad}/(\gamma_{rad} + \gamma_{sc})$ vs frequency for different electron Fermi energies in the layers. Dashed lines are just a guide for eye}
    \label{fig:RadDecayProbability}
\end{figure}

Finally, we estimate the rate of plasmon absorption due to the Drude loss. It is worthwhile noting that $\gamma_{abs}$ is not just the inverse of electron momentum relaxation rate due to the non-negligible spatial dispersion of conductivity~\cite{Principi_plasmon_loss_hBN}. To account for the spatial dispersion and electron scattering simultaneously, one can solve the kinetic equation for electrons with particle-conserving collision integral~\cite{Plasmons_Coupled}. This leads to the modification of collisionless conductivity (\ref{InPlane-cond}) according to
\begin{equation}
     \sigma = i \frac{g}{2\pi} \frac{e^2}{\hbar} \frac{\varepsilon_F}{q v} x \left\{ \frac{x/\sqrt{x^2-1} - 1}{1 - (i\nu/\omega) [x/\sqrt{x^2-1} - 1]} \right\},
\end{equation}
where $x= (\omega + i \nu)/qv$, and $\nu$ is the electron scattering rate. Considering electron-phonon and electron-impurity collisions as the dominant scattering sources, we evaluate $\nu$ as~\cite{Vasko-Ryzhii,DasSarma_Impurity}
\begin{equation}
     \nu = \frac{\varepsilon_F}{T}\frac{D^2 T^2}{4\rho c_s^2 v^2} + \frac{\pi}{16} \frac{v^2 N_i}{\varepsilon_F} J(\alpha_c).
\end{equation}
Here $D\approx 30$ eV is the deformation potential in graphene, $\rho = 7.6\times 10^{-7}$ kg/m$^2$ is its mass density, and $c_s =2 \times 10^4$ m/s is the sound velocity, $N_i$ is the impurity density, and $J(\alpha_c)$ is the dimensionless integral
\begin{equation}
     J(\alpha_c) = \int\limits_{0}^{2\pi}\frac{d\theta (1-\cos^2\theta)}{[1+(2\alpha_c)^{-1} \sin(\theta/2)]^2}.
\end{equation}
The results of scattering rate and conversion efficiency calculation are shown in Figs. (\ref{fig:RadvsScattering}) and (\ref{fig:RadDecayProbability}), respectively. An increase in scattering rate with reducing the Fermi energy in Fig.~(\ref{fig:RadvsScattering}) is due to the impurity scattering contribution to plasmon damping which scales as $\varepsilon_F^{-1}$. At lower impurity density, the scattering will be dominated by phonons and an increase in Fermi energy would increase the scattering rates.

\bibliography{Bibliography}

\begin{thebibliography}{48}%
\makeatletter
\providecommand \@ifxundefined [1]{%
 \@ifx{#1\undefined}
}%
\providecommand \@ifnum [1]{%
 \ifnum #1\expandafter \@firstoftwo
 \else \expandafter \@secondoftwo
 \fi
}%
\providecommand \@ifx [1]{%
 \ifx #1\expandafter \@firstoftwo
 \else \expandafter \@secondoftwo
 \fi
}%
\providecommand \natexlab [1]{#1}%
\providecommand \enquote  [1]{``#1''}%
\providecommand \bibnamefont  [1]{#1}%
\providecommand \bibfnamefont [1]{#1}%
\providecommand \citenamefont [1]{#1}%
\providecommand \href@noop [0]{\@secondoftwo}%
\providecommand \href [0]{\begingroup \@sanitize@url \@href}%
\providecommand \@href[1]{\@@startlink{#1}\@@href}%
\providecommand \@@href[1]{\endgroup#1\@@endlink}%
\providecommand \@sanitize@url [0]{\catcode `\\12\catcode `\$12\catcode
  `\&12\catcode `\#12\catcode `\^12\catcode `\_12\catcode `\%12\relax}%
\providecommand \@@startlink[1]{}%
\providecommand \@@endlink[0]{}%
\providecommand \url  [0]{\begingroup\@sanitize@url \@url }%
\providecommand \@url [1]{\endgroup\@href {#1}{\urlprefix }}%
\providecommand \urlprefix  [0]{URL }%
\providecommand \Eprint [0]{\href }%
\providecommand \doibase [0]{http://dx.doi.org/}%
\providecommand \selectlanguage [0]{\@gobble}%
\providecommand \bibinfo  [0]{\@secondoftwo}%
\providecommand \bibfield  [0]{\@secondoftwo}%
\providecommand \translation [1]{[#1]}%
\providecommand \BibitemOpen [0]{}%
\providecommand \bibitemStop [0]{}%
\providecommand \bibitemNoStop [0]{.\EOS\space}%
\providecommand \EOS [0]{\spacefactor3000\relax}%
\providecommand \BibitemShut  [1]{\csname bibitem#1\endcsname}%
\let\auto@bib@innerbib\@empty
\bibitem [{\citenamefont {Woessner}\ \emph {et~al.}(2015)\citenamefont
  {Woessner}, \citenamefont {Lundeberg}, \citenamefont {Gao}, \citenamefont
  {Principi}, \citenamefont {Alonso-Gonz{\'a}lez}, \citenamefont {Carrega},
  \citenamefont {Watanabe}, \citenamefont {Taniguchi}, \citenamefont {Vignale},
  \citenamefont {Polini}, \citenamefont {Hone}, \citenamefont {Hillenbrand},\
  and\ \citenamefont {Koppens}}]{Koppens_nano_imaging_hBN}%
  \BibitemOpen
  \bibfield  {author} {\bibinfo {author} {\bibfnamefont {A.}~\bibnamefont
  {Woessner}}, \bibinfo {author} {\bibfnamefont {M.~B.}\ \bibnamefont
  {Lundeberg}}, \bibinfo {author} {\bibfnamefont {Y.}~\bibnamefont {Gao}},
  \bibinfo {author} {\bibfnamefont {A.}~\bibnamefont {Principi}}, \bibinfo
  {author} {\bibfnamefont {P.}~\bibnamefont {Alonso-Gonz{\'a}lez}}, \bibinfo
  {author} {\bibfnamefont {M.}~\bibnamefont {Carrega}}, \bibinfo {author}
  {\bibfnamefont {K.}~\bibnamefont {Watanabe}}, \bibinfo {author}
  {\bibfnamefont {T.}~\bibnamefont {Taniguchi}}, \bibinfo {author}
  {\bibfnamefont {G.}~\bibnamefont {Vignale}}, \bibinfo {author} {\bibfnamefont
  {M.}~\bibnamefont {Polini}}, \bibinfo {author} {\bibfnamefont
  {J.}~\bibnamefont {Hone}}, \bibinfo {author} {\bibfnamefont {R.}~\bibnamefont
  {Hillenbrand}}, \ and\ \bibinfo {author} {\bibfnamefont {F.~H.~L.}\
  \bibnamefont {Koppens}},\ }\href@noop {} {\bibfield  {journal} {\bibinfo
  {journal} {Nat. Mater.}\ }\textbf {\bibinfo {volume} {14}},\ \bibinfo {pages}
  {421} (\bibinfo {year} {2015})}\BibitemShut {NoStop}%
\bibitem [{\citenamefont {Koppens}\ \emph {et~al.}(2011)\citenamefont
  {Koppens}, \citenamefont {Chang},\ and\ \citenamefont {Garcia~de
  Abajo}}]{Graphene_plasmonics-2}%
  \BibitemOpen
  \bibfield  {author} {\bibinfo {author} {\bibfnamefont {F.~H.~L.}\
  \bibnamefont {Koppens}}, \bibinfo {author} {\bibfnamefont {D.~E.}\
  \bibnamefont {Chang}}, \ and\ \bibinfo {author} {\bibfnamefont {F.~J.}\
  \bibnamefont {Garcia~de Abajo}},\ }\href {\doibase 10.1021/nl201771h}
  {\bibfield  {journal} {\bibinfo  {journal} {Nano Lett.}\ }\textbf {\bibinfo
  {volume} {11}},\ \bibinfo {pages} {3370} (\bibinfo {year}
  {2011})}\BibitemShut {NoStop}%
\bibitem [{\citenamefont {Bostwick}\ \emph {et~al.}(2010)\citenamefont
  {Bostwick}, \citenamefont {Speck}, \citenamefont {Seyller}, \citenamefont
  {Horn}, \citenamefont {Polini}, \citenamefont {Asgari}, \citenamefont
  {MacDonald},\ and\ \citenamefont {Rotenberg}}]{Bostwick}%
  \BibitemOpen
  \bibfield  {author} {\bibinfo {author} {\bibfnamefont {A.}~\bibnamefont
  {Bostwick}}, \bibinfo {author} {\bibfnamefont {F.}~\bibnamefont {Speck}},
  \bibinfo {author} {\bibfnamefont {T.}~\bibnamefont {Seyller}}, \bibinfo
  {author} {\bibfnamefont {K.}~\bibnamefont {Horn}}, \bibinfo {author}
  {\bibfnamefont {M.}~\bibnamefont {Polini}}, \bibinfo {author} {\bibfnamefont
  {R.}~\bibnamefont {Asgari}}, \bibinfo {author} {\bibfnamefont {A.~H.}\
  \bibnamefont {MacDonald}}, \ and\ \bibinfo {author} {\bibfnamefont
  {E.}~\bibnamefont {Rotenberg}},\ }\href {\doibase 10.1126/science.1186489}
  {\bibfield  {journal} {\bibinfo  {journal} {Science}\ }\textbf {\bibinfo
  {volume} {328}},\ \bibinfo {pages} {999} (\bibinfo {year}
  {2010})}\BibitemShut {NoStop}%
\bibitem [{\citenamefont {Thongrattanasiri}\ \emph {et~al.}(2012)\citenamefont
  {Thongrattanasiri}, \citenamefont {Koppens},\ and\ \citenamefont {Garcia~de
  Abajo}}]{Perfect_absorption}%
  \BibitemOpen
  \bibfield  {author} {\bibinfo {author} {\bibfnamefont {S.}~\bibnamefont
  {Thongrattanasiri}}, \bibinfo {author} {\bibfnamefont {F.~H.~L.}\
  \bibnamefont {Koppens}}, \ and\ \bibinfo {author} {\bibfnamefont {F.~J.}\
  \bibnamefont {Garcia~de Abajo}},\ }\href {\doibase
  10.1103/PhysRevLett.108.047401} {\bibfield  {journal} {\bibinfo  {journal}
  {Phys. Rev. Lett.}\ }\textbf {\bibinfo {volume} {108}},\ \bibinfo {pages}
  {047401} (\bibinfo {year} {2012})}\BibitemShut {NoStop}%
\bibitem [{\citenamefont {Rana}(2008)}]{Rana_IEEE}%
  \BibitemOpen
  \bibfield  {author} {\bibinfo {author} {\bibfnamefont {F.}~\bibnamefont
  {Rana}},\ }\href {\doibase 10.1109/TNANO.2007.910334} {\bibfield  {journal}
  {\bibinfo  {journal} {IEEE T. Nanotechnol.}\ }\textbf {\bibinfo {volume}
  {7}},\ \bibinfo {pages} {91} (\bibinfo {year} {2008})}\BibitemShut {NoStop}%
\bibitem [{\citenamefont {Watanabe}\ \emph {et~al.}(2013)\citenamefont
  {Watanabe}, \citenamefont {Fukushima}, \citenamefont {Yabe}, \citenamefont
  {Tombet}, \citenamefont {Satou}, \citenamefont {Dubinov}, \citenamefont
  {Aleshkin}, \citenamefont {Mitin}, \citenamefont {Ryzhii},\ and\
  \citenamefont {Otsuji}}]{Watanabe_gain_enhancement}%
  \BibitemOpen
  \bibfield  {author} {\bibinfo {author} {\bibfnamefont {T.}~\bibnamefont
  {Watanabe}}, \bibinfo {author} {\bibfnamefont {T.}~\bibnamefont {Fukushima}},
  \bibinfo {author} {\bibfnamefont {Y.}~\bibnamefont {Yabe}}, \bibinfo {author}
  {\bibfnamefont {S.~A.~B.}\ \bibnamefont {Tombet}}, \bibinfo {author}
  {\bibfnamefont {A.}~\bibnamefont {Satou}}, \bibinfo {author} {\bibfnamefont
  {A.~A.}\ \bibnamefont {Dubinov}}, \bibinfo {author} {\bibfnamefont {V.~Y.}\
  \bibnamefont {Aleshkin}}, \bibinfo {author} {\bibfnamefont {V.}~\bibnamefont
  {Mitin}}, \bibinfo {author} {\bibfnamefont {V.}~\bibnamefont {Ryzhii}}, \
  and\ \bibinfo {author} {\bibfnamefont {T.}~\bibnamefont {Otsuji}},\ }\href
  {http://stacks.iop.org/1367-2630/15/i=7/a=075003} {\bibfield  {journal}
  {\bibinfo  {journal} {New Journal of Physics}\ }\textbf {\bibinfo {volume}
  {15}},\ \bibinfo {pages} {075003} (\bibinfo {year} {2013})}\BibitemShut
  {NoStop}%
\bibitem [{\citenamefont {Ryzhii}\ \emph {et~al.}(2007)\citenamefont {Ryzhii},
  \citenamefont {Satou},\ and\ \citenamefont {Otsuji}}]{Ryzhii-plasmons}%
  \BibitemOpen
  \bibfield  {author} {\bibinfo {author} {\bibfnamefont {V.}~\bibnamefont
  {Ryzhii}}, \bibinfo {author} {\bibfnamefont {A.}~\bibnamefont {Satou}}, \
  and\ \bibinfo {author} {\bibfnamefont {T.}~\bibnamefont {Otsuji}},\ }\href
  {\doibase 10.1063/1.2426904} {\bibfield  {journal} {\bibinfo  {journal} {J.
  Appl. Phys.}\ }\textbf {\bibinfo {volume} {101}},\ \bibinfo {eid} {024509}
  (\bibinfo {year} {2007})}\BibitemShut {NoStop}%
\bibitem [{\citenamefont {Alonso-Gonzalez}\ \emph {et~al.}(2017)\citenamefont
  {Alonso-Gonzalez}, \citenamefont {Nikitin}, \citenamefont {Gao},
  \citenamefont {Woessner}, \citenamefont {Lundeberg}, \citenamefont
  {Principi}, \citenamefont {Forcellini}, \citenamefont {Yan}, \citenamefont
  {Velez}, \citenamefont {Huber}, \citenamefont {Watanabe}, \citenamefont
  {Taniguchi}, \citenamefont {Casanova}, \citenamefont {E.~Hueso},
  \citenamefont {Polini}, \citenamefont {Hone}, \citenamefont {Koppens},\ and\
  \citenamefont {Hillenbrand}}]{Koppens_AcousticPlasmons}%
  \BibitemOpen
  \bibfield  {author} {\bibinfo {author} {\bibfnamefont {P.}~\bibnamefont
  {Alonso-Gonzalez}}, \bibinfo {author} {\bibfnamefont {A.~Y.}\ \bibnamefont
  {Nikitin}}, \bibinfo {author} {\bibfnamefont {Y.}~\bibnamefont {Gao}},
  \bibinfo {author} {\bibfnamefont {A.}~\bibnamefont {Woessner}}, \bibinfo
  {author} {\bibfnamefont {M.~B.}\ \bibnamefont {Lundeberg}}, \bibinfo {author}
  {\bibfnamefont {A.}~\bibnamefont {Principi}}, \bibinfo {author}
  {\bibfnamefont {N.}~\bibnamefont {Forcellini}}, \bibinfo {author}
  {\bibfnamefont {W.}~\bibnamefont {Yan}}, \bibinfo {author} {\bibfnamefont
  {S.}~\bibnamefont {Velez}}, \bibinfo {author} {\bibfnamefont
  {A.}~\bibnamefont {Huber}}, \bibinfo {author} {\bibfnamefont
  {K.}~\bibnamefont {Watanabe}}, \bibinfo {author} {\bibfnamefont
  {T.}~\bibnamefont {Taniguchi}}, \bibinfo {author} {\bibfnamefont
  {F.}~\bibnamefont {Casanova}}, \bibinfo {author} {\bibfnamefont
  {L.}~\bibnamefont {E.~Hueso}}, \bibinfo {author} {\bibfnamefont
  {M.}~\bibnamefont {Polini}}, \bibinfo {author} {\bibfnamefont
  {J.}~\bibnamefont {Hone}}, \bibinfo {author} {\bibfnamefont {F.~H.~L.}\
  \bibnamefont {Koppens}}, \ and\ \bibinfo {author} {\bibfnamefont
  {R.}~\bibnamefont {Hillenbrand}},\ }\href@noop {} {\bibfield  {journal}
  {\bibinfo  {journal} {Nature Nanotechnology}\ }\textbf {\bibinfo {volume}
  {12}},\ \bibinfo {pages} {31–35} (\bibinfo {year} {2017})}\BibitemShut
  {NoStop}%
\bibitem [{\citenamefont {Gaskell}\ \emph {et~al.}(2015)\citenamefont
  {Gaskell}, \citenamefont {Eaves}, \citenamefont {Novoselov}, \citenamefont
  {Mishchenko}, \citenamefont {Geim}, \citenamefont {Fromhold},\ and\
  \citenamefont {Greenaway}}]{Novoselov_APL_RTD}%
  \BibitemOpen
  \bibfield  {author} {\bibinfo {author} {\bibfnamefont {J.}~\bibnamefont
  {Gaskell}}, \bibinfo {author} {\bibfnamefont {L.}~\bibnamefont {Eaves}},
  \bibinfo {author} {\bibfnamefont {K.~S.}\ \bibnamefont {Novoselov}}, \bibinfo
  {author} {\bibfnamefont {A.}~\bibnamefont {Mishchenko}}, \bibinfo {author}
  {\bibfnamefont {A.~K.}\ \bibnamefont {Geim}}, \bibinfo {author}
  {\bibfnamefont {T.~M.}\ \bibnamefont {Fromhold}}, \ and\ \bibinfo {author}
  {\bibfnamefont {M.~T.}\ \bibnamefont {Greenaway}},\ }\href {\doibase
  10.1063/1.4930230} {\bibfield  {journal} {\bibinfo  {journal} {Applied
  Physics Letters}\ }\textbf {\bibinfo {volume} {107}},\ \bibinfo {pages}
  {103105} (\bibinfo {year} {2015})}\BibitemShut {NoStop}%
\bibitem [{\citenamefont {Britnell}\ \emph {et~al.}(2013)\citenamefont
  {Britnell}, \citenamefont {Gorbachev}, \citenamefont {Geim}, \citenamefont
  {Ponomarenko}, \citenamefont {Mishchenko}, \citenamefont {Greenaway},
  \citenamefont {Fromhold}, \citenamefont {Novoselov},\ and\ \citenamefont
  {Eaves}}]{britnell2013resonant}%
  \BibitemOpen
  \bibfield  {author} {\bibinfo {author} {\bibfnamefont {L.}~\bibnamefont
  {Britnell}}, \bibinfo {author} {\bibfnamefont {R.}~\bibnamefont {Gorbachev}},
  \bibinfo {author} {\bibfnamefont {A.}~\bibnamefont {Geim}}, \bibinfo {author}
  {\bibfnamefont {L.}~\bibnamefont {Ponomarenko}}, \bibinfo {author}
  {\bibfnamefont {A.}~\bibnamefont {Mishchenko}}, \bibinfo {author}
  {\bibfnamefont {M.}~\bibnamefont {Greenaway}}, \bibinfo {author}
  {\bibfnamefont {T.}~\bibnamefont {Fromhold}}, \bibinfo {author}
  {\bibfnamefont {K.}~\bibnamefont {Novoselov}}, \ and\ \bibinfo {author}
  {\bibfnamefont {L.}~\bibnamefont {Eaves}},\ }\href@noop {} {\bibfield
  {journal} {\bibinfo  {journal} {Nat. Communications}\ }\textbf {\bibinfo
  {volume} {4}},\ \bibinfo {pages} {1794} (\bibinfo {year} {2013})}\BibitemShut
  {NoStop}%
\bibitem [{\citenamefont {Greenaway}\ \emph {et~al.}(2015)\citenamefont
  {Greenaway}, \citenamefont {Vdovin}, \citenamefont {Mishchenko},
  \citenamefont {Makarovsky}, \citenamefont {Patan{\`e}}, \citenamefont
  {Wallbank}, \citenamefont {Cao}, \citenamefont {Kretinin}, \citenamefont
  {Zhu}, \citenamefont {Morozov}, \citenamefont {Fal’ko}, \citenamefont
  {Novoselov}, \citenamefont {Geim}, \citenamefont {Fromhold},\ and\
  \citenamefont {Eaves}}]{Chiral_tunneling}%
  \BibitemOpen
  \bibfield  {author} {\bibinfo {author} {\bibfnamefont {M.}~\bibnamefont
  {Greenaway}}, \bibinfo {author} {\bibfnamefont {E.}~\bibnamefont {Vdovin}},
  \bibinfo {author} {\bibfnamefont {A.}~\bibnamefont {Mishchenko}}, \bibinfo
  {author} {\bibfnamefont {O.}~\bibnamefont {Makarovsky}}, \bibinfo {author}
  {\bibfnamefont {A.}~\bibnamefont {Patan{\`e}}}, \bibinfo {author}
  {\bibfnamefont {J.}~\bibnamefont {Wallbank}}, \bibinfo {author}
  {\bibfnamefont {Y.}~\bibnamefont {Cao}}, \bibinfo {author} {\bibfnamefont
  {A.}~\bibnamefont {Kretinin}}, \bibinfo {author} {\bibfnamefont
  {M.}~\bibnamefont {Zhu}}, \bibinfo {author} {\bibfnamefont {S.}~\bibnamefont
  {Morozov}}, \bibinfo {author} {\bibfnamefont {V.~I.}\ \bibnamefont
  {Fal’ko}}, \bibinfo {author} {\bibfnamefont {K.}~\bibnamefont {Novoselov}},
  \bibinfo {author} {\bibfnamefont {A.}~\bibnamefont {Geim}}, \bibinfo {author}
  {\bibfnamefont {T.}~\bibnamefont {Fromhold}}, \ and\ \bibinfo {author}
  {\bibfnamefont {L.}~\bibnamefont {Eaves}},\ }\href@noop {} {\bibfield
  {journal} {\bibinfo  {journal} {Nature Physics}\ }\textbf {\bibinfo {volume}
  {11}},\ \bibinfo {pages} {1057} (\bibinfo {year} {2015})}\BibitemShut
  {NoStop}%
\bibitem [{\citenamefont {Wallbank}\ \emph {et~al.}(2016)\citenamefont
  {Wallbank}, \citenamefont {Ghazaryan}, \citenamefont {Misra}, \citenamefont
  {Cao}, \citenamefont {Tu}, \citenamefont {Piot}, \citenamefont {Potemski},
  \citenamefont {Pezzini}, \citenamefont {Wiedmann}, \citenamefont {Zeitler},
  \citenamefont {Lane}, \citenamefont {Morozov}, \citenamefont {Greenaway},
  \citenamefont {Eaves}, \citenamefont {Geim}, \citenamefont {Fal'ko},
  \citenamefont {Novoselov},\ and\ \citenamefont
  {Mishchenko}}]{Novoselov_Magnetic_tunneling}%
  \BibitemOpen
  \bibfield  {author} {\bibinfo {author} {\bibfnamefont {J.~R.}\ \bibnamefont
  {Wallbank}}, \bibinfo {author} {\bibfnamefont {D.}~\bibnamefont {Ghazaryan}},
  \bibinfo {author} {\bibfnamefont {A.}~\bibnamefont {Misra}}, \bibinfo
  {author} {\bibfnamefont {Y.}~\bibnamefont {Cao}}, \bibinfo {author}
  {\bibfnamefont {J.~S.}\ \bibnamefont {Tu}}, \bibinfo {author} {\bibfnamefont
  {B.~A.}\ \bibnamefont {Piot}}, \bibinfo {author} {\bibfnamefont
  {M.}~\bibnamefont {Potemski}}, \bibinfo {author} {\bibfnamefont
  {S.}~\bibnamefont {Pezzini}}, \bibinfo {author} {\bibfnamefont
  {S.}~\bibnamefont {Wiedmann}}, \bibinfo {author} {\bibfnamefont
  {U.}~\bibnamefont {Zeitler}}, \bibinfo {author} {\bibfnamefont {T.~L.~M.}\
  \bibnamefont {Lane}}, \bibinfo {author} {\bibfnamefont {S.~V.}\ \bibnamefont
  {Morozov}}, \bibinfo {author} {\bibfnamefont {M.~T.}\ \bibnamefont
  {Greenaway}}, \bibinfo {author} {\bibfnamefont {L.}~\bibnamefont {Eaves}},
  \bibinfo {author} {\bibfnamefont {A.~K.}\ \bibnamefont {Geim}}, \bibinfo
  {author} {\bibfnamefont {V.~I.}\ \bibnamefont {Fal'ko}}, \bibinfo {author}
  {\bibfnamefont {K.~S.}\ \bibnamefont {Novoselov}}, \ and\ \bibinfo {author}
  {\bibfnamefont {A.}~\bibnamefont {Mishchenko}},\ }\href {\doibase
  10.1126/science.aaf4621} {\bibfield  {journal} {\bibinfo  {journal}
  {Science}\ }\textbf {\bibinfo {volume} {353}},\ \bibinfo {pages} {575}
  (\bibinfo {year} {2016})}\BibitemShut {NoStop}%
\bibitem [{\citenamefont {Vdovin}\ \emph {et~al.}(2016)\citenamefont {Vdovin},
  \citenamefont {Mishchenko}, \citenamefont {Greenaway}, \citenamefont {Zhu},
  \citenamefont {Ghazaryan}, \citenamefont {Misra}, \citenamefont {Cao},
  \citenamefont {Morozov}, \citenamefont {Makarovsky}, \citenamefont
  {Fromhold}, \citenamefont {Patan\`e}, \citenamefont {Slotman}, \citenamefont
  {Katsnelson}, \citenamefont {Geim}, \citenamefont {Novoselov},\ and\
  \citenamefont {Eaves}}]{Phonon_assisted_graphene}%
  \BibitemOpen
  \bibfield  {author} {\bibinfo {author} {\bibfnamefont {E.~E.}\ \bibnamefont
  {Vdovin}}, \bibinfo {author} {\bibfnamefont {A.}~\bibnamefont {Mishchenko}},
  \bibinfo {author} {\bibfnamefont {M.~T.}\ \bibnamefont {Greenaway}}, \bibinfo
  {author} {\bibfnamefont {M.~J.}\ \bibnamefont {Zhu}}, \bibinfo {author}
  {\bibfnamefont {D.}~\bibnamefont {Ghazaryan}}, \bibinfo {author}
  {\bibfnamefont {A.}~\bibnamefont {Misra}}, \bibinfo {author} {\bibfnamefont
  {Y.}~\bibnamefont {Cao}}, \bibinfo {author} {\bibfnamefont {S.~V.}\
  \bibnamefont {Morozov}}, \bibinfo {author} {\bibfnamefont {O.}~\bibnamefont
  {Makarovsky}}, \bibinfo {author} {\bibfnamefont {T.~M.}\ \bibnamefont
  {Fromhold}}, \bibinfo {author} {\bibfnamefont {A.}~\bibnamefont {Patan\`e}},
  \bibinfo {author} {\bibfnamefont {G.~J.}\ \bibnamefont {Slotman}}, \bibinfo
  {author} {\bibfnamefont {M.~I.}\ \bibnamefont {Katsnelson}}, \bibinfo
  {author} {\bibfnamefont {A.~K.}\ \bibnamefont {Geim}}, \bibinfo {author}
  {\bibfnamefont {K.~S.}\ \bibnamefont {Novoselov}}, \ and\ \bibinfo {author}
  {\bibfnamefont {L.}~\bibnamefont {Eaves}},\ }\href {\doibase
  10.1103/PhysRevLett.116.186603} {\bibfield  {journal} {\bibinfo  {journal}
  {Phys. Rev. Lett.}\ }\textbf {\bibinfo {volume} {116}},\ \bibinfo {pages}
  {186603} (\bibinfo {year} {2016})}\BibitemShut {NoStop}%
\bibitem [{\citenamefont {Esaki}\ \emph {et~al.}(1968)\citenamefont {Esaki},
  \citenamefont {Chang}, \citenamefont {Stiles}, \citenamefont {O'Kane},\ and\
  \citenamefont {Wiser}}]{Esaki_Phonon}%
  \BibitemOpen
  \bibfield  {author} {\bibinfo {author} {\bibfnamefont {L.}~\bibnamefont
  {Esaki}}, \bibinfo {author} {\bibfnamefont {L.~L.}\ \bibnamefont {Chang}},
  \bibinfo {author} {\bibfnamefont {P.~J.}\ \bibnamefont {Stiles}}, \bibinfo
  {author} {\bibfnamefont {D.~F.}\ \bibnamefont {O'Kane}}, \ and\ \bibinfo
  {author} {\bibfnamefont {N.}~\bibnamefont {Wiser}},\ }\href {\doibase
  10.1103/PhysRev.167.637} {\bibfield  {journal} {\bibinfo  {journal} {Phys.
  Rev.}\ }\textbf {\bibinfo {volume} {167}},\ \bibinfo {pages} {637} (\bibinfo
  {year} {1968})}\BibitemShut {NoStop}%
\bibitem [{\citenamefont {Zhang}\ \emph {et~al.}(1994)\citenamefont {Zhang},
  \citenamefont {Lerch}, \citenamefont {Martin}, \citenamefont {Simmonds},\
  and\ \citenamefont {Eaves}}]{Plasmon-assisted_Eaves}%
  \BibitemOpen
  \bibfield  {author} {\bibinfo {author} {\bibfnamefont {C.}~\bibnamefont
  {Zhang}}, \bibinfo {author} {\bibfnamefont {M.~L.~F.}\ \bibnamefont {Lerch}},
  \bibinfo {author} {\bibfnamefont {A.~D.}\ \bibnamefont {Martin}}, \bibinfo
  {author} {\bibfnamefont {P.~E.}\ \bibnamefont {Simmonds}}, \ and\ \bibinfo
  {author} {\bibfnamefont {L.}~\bibnamefont {Eaves}},\ }\href {\doibase
  10.1103/PhysRevLett.72.3397} {\bibfield  {journal} {\bibinfo  {journal}
  {Phys. Rev. Lett.}\ }\textbf {\bibinfo {volume} {72}},\ \bibinfo {pages}
  {3397} (\bibinfo {year} {1994})}\BibitemShut {NoStop}%
\bibitem [{\citenamefont {Belenov}\ \emph {et~al.}(1987)\citenamefont
  {Belenov}, \citenamefont {Luskinovich}, \citenamefont {Romanenko},
  \citenamefont {Sobolev},\ and\ \citenamefont
  {Uskov}}]{Belenov_Emission_resonant}%
  \BibitemOpen
  \bibfield  {author} {\bibinfo {author} {\bibfnamefont {E.~M.}\ \bibnamefont
  {Belenov}}, \bibinfo {author} {\bibfnamefont {P.~N.}\ \bibnamefont
  {Luskinovich}}, \bibinfo {author} {\bibfnamefont {V.~I.}\ \bibnamefont
  {Romanenko}}, \bibinfo {author} {\bibfnamefont {A.~G.}\ \bibnamefont
  {Sobolev}}, \ and\ \bibinfo {author} {\bibfnamefont {A.~V.}\ \bibnamefont
  {Uskov}},\ }\href {http://stacks.iop.org/0049-1748/17/i=10/a=A33} {\bibfield
  {journal} {\bibinfo  {journal} {Soviet Journal of Quantum Electronics}\
  }\textbf {\bibinfo {volume} {17}},\ \bibinfo {pages} {1348} (\bibinfo {year}
  {1987})}\BibitemShut {NoStop}%
\bibitem [{\citenamefont {Kempa}\ \emph {et~al.}(2001)\citenamefont {Kempa},
  \citenamefont {Gornik}, \citenamefont {Unterrainer}, \citenamefont {Kast},\
  and\ \citenamefont {Strasser}}]{Kempa_PlasmonAssisted}%
  \BibitemOpen
  \bibfield  {author} {\bibinfo {author} {\bibfnamefont {K.}~\bibnamefont
  {Kempa}}, \bibinfo {author} {\bibfnamefont {E.}~\bibnamefont {Gornik}},
  \bibinfo {author} {\bibfnamefont {K.}~\bibnamefont {Unterrainer}}, \bibinfo
  {author} {\bibfnamefont {M.}~\bibnamefont {Kast}}, \ and\ \bibinfo {author}
  {\bibfnamefont {G.}~\bibnamefont {Strasser}},\ }\href {\doibase
  10.1103/PhysRevLett.86.2850} {\bibfield  {journal} {\bibinfo  {journal}
  {Phys. Rev. Lett.}\ }\textbf {\bibinfo {volume} {86}},\ \bibinfo {pages}
  {2850} (\bibinfo {year} {2001})}\BibitemShut {NoStop}%
\bibitem [{\citenamefont {Lambe}\ and\ \citenamefont
  {McCarthy}(1976)}]{Lambe_PRL}%
  \BibitemOpen
  \bibfield  {author} {\bibinfo {author} {\bibfnamefont {J.}~\bibnamefont
  {Lambe}}\ and\ \bibinfo {author} {\bibfnamefont {S.~L.}\ \bibnamefont
  {McCarthy}},\ }\href {\doibase 10.1103/PhysRevLett.37.923} {\bibfield
  {journal} {\bibinfo  {journal} {Phys. Rev. Lett.}\ }\textbf {\bibinfo
  {volume} {37}},\ \bibinfo {pages} {923} (\bibinfo {year} {1976})}\BibitemShut
  {NoStop}%
\bibitem [{\citenamefont {Parzefall}\ \emph {et~al.}(2017)\citenamefont
  {Parzefall}, \citenamefont {Bharadwaj},\ and\ \citenamefont
  {Novotny}}]{Parzefall2017}%
  \BibitemOpen
  \bibfield  {author} {\bibinfo {author} {\bibfnamefont {M.}~\bibnamefont
  {Parzefall}}, \bibinfo {author} {\bibfnamefont {P.}~\bibnamefont
  {Bharadwaj}}, \ and\ \bibinfo {author} {\bibfnamefont {L.}~\bibnamefont
  {Novotny}},\ }\enquote {\bibinfo {title} {Antenna-coupled tunnel
  junctions},}\ in\ \href {\doibase 10.1007/978-3-319-45820-5_10} {\emph
  {\bibinfo {booktitle} {Quantum Plasmonics}}},\ \bibinfo {editor} {edited by\
  \bibinfo {editor} {\bibfnamefont {S.~I.}\ \bibnamefont {Bozhevolnyi}},
  \bibinfo {editor} {\bibfnamefont {L.}~\bibnamefont {Martin-Moreno}}, \ and\
  \bibinfo {editor} {\bibfnamefont {F.}~\bibnamefont {Garcia-Vidal}}}\
  (\bibinfo  {publisher} {Springer International Publishing},\ \bibinfo
  {address} {Cham},\ \bibinfo {year} {2017})\ pp.\ \bibinfo {pages}
  {211--236}\BibitemShut {NoStop}%
\bibitem [{\citenamefont {Yadav}\ \emph {et~al.}(2016)\citenamefont {Yadav},
  \citenamefont {Tombet}, \citenamefont {Watanabe}, \citenamefont {Arnold},
  \citenamefont {Ryzhii},\ and\ \citenamefont {Otsuji}}]{Yadav_THz}%
  \BibitemOpen
  \bibfield  {author} {\bibinfo {author} {\bibfnamefont {D.}~\bibnamefont
  {Yadav}}, \bibinfo {author} {\bibfnamefont {S.~B.}\ \bibnamefont {Tombet}},
  \bibinfo {author} {\bibfnamefont {T.}~\bibnamefont {Watanabe}}, \bibinfo
  {author} {\bibfnamefont {S.}~\bibnamefont {Arnold}}, \bibinfo {author}
  {\bibfnamefont {V.}~\bibnamefont {Ryzhii}}, \ and\ \bibinfo {author}
  {\bibfnamefont {T.}~\bibnamefont {Otsuji}},\ }\href
  {http://stacks.iop.org/2053-1583/3/i=4/a=045009} {\bibfield  {journal}
  {\bibinfo  {journal} {2D Materials}\ }\textbf {\bibinfo {volume} {3}},\
  \bibinfo {pages} {045009} (\bibinfo {year} {2016})}\BibitemShut {NoStop}%
\bibitem [{\citenamefont {Hwang}\ and\ \citenamefont
  {Das~Sarma}(2009)}]{Hwang_PRB_2GL}%
  \BibitemOpen
  \bibfield  {author} {\bibinfo {author} {\bibfnamefont {E.~H.}\ \bibnamefont
  {Hwang}}\ and\ \bibinfo {author} {\bibfnamefont {S.}~\bibnamefont
  {Das~Sarma}},\ }\href {\doibase 10.1103/PhysRevB.80.205405} {\bibfield
  {journal} {\bibinfo  {journal} {Phys. Rev. B}\ }\textbf {\bibinfo {volume}
  {80}},\ \bibinfo {pages} {205405} (\bibinfo {year} {2009})}\BibitemShut
  {NoStop}%
\bibitem [{\citenamefont {Svintsov}\ \emph {et~al.}(2013)\citenamefont
  {Svintsov}, \citenamefont {Vyurkov}, \citenamefont {Ryzhii},\ and\
  \citenamefont {Otsuji}}]{Voltage_controlled}%
  \BibitemOpen
  \bibfield  {author} {\bibinfo {author} {\bibfnamefont {D.}~\bibnamefont
  {Svintsov}}, \bibinfo {author} {\bibfnamefont {V.}~\bibnamefont {Vyurkov}},
  \bibinfo {author} {\bibfnamefont {V.}~\bibnamefont {Ryzhii}}, \ and\ \bibinfo
  {author} {\bibfnamefont {T.}~\bibnamefont {Otsuji}},\ }\href {\doibase
  http://dx.doi.org/10.1063/1.4789818} {\bibfield  {journal} {\bibinfo
  {journal} {J. Appl. Phys.}\ }\textbf {\bibinfo {volume} {113}},\ \bibinfo
  {eid} {053701} (\bibinfo {year} {2013})}\BibitemShut {NoStop}%
\bibitem [{\citenamefont {Svintsov}\ \emph {et~al.}(2016)\citenamefont
  {Svintsov}, \citenamefont {Devizorova}, \citenamefont {Otsuji},\ and\
  \citenamefont {Ryzhii}}]{Plasmons_Coupled}%
  \BibitemOpen
  \bibfield  {author} {\bibinfo {author} {\bibfnamefont {D.}~\bibnamefont
  {Svintsov}}, \bibinfo {author} {\bibfnamefont {Z.}~\bibnamefont
  {Devizorova}}, \bibinfo {author} {\bibfnamefont {T.}~\bibnamefont {Otsuji}},
  \ and\ \bibinfo {author} {\bibfnamefont {V.}~\bibnamefont {Ryzhii}},\ }\href
  {\doibase 10.1103/PhysRevB.94.115301} {\bibfield  {journal} {\bibinfo
  {journal} {Phys. Rev. B}\ }\textbf {\bibinfo {volume} {94}},\ \bibinfo
  {pages} {115301} (\bibinfo {year} {2016})}\BibitemShut {NoStop}%
\bibitem [{\citenamefont {Sensale-Rodriguez}(2013)}]{Berardi_APL}%
  \BibitemOpen
  \bibfield  {author} {\bibinfo {author} {\bibfnamefont {B.}~\bibnamefont
  {Sensale-Rodriguez}},\ }\href {\doibase http://dx.doi.org/10.1063/1.4821221}
  {\bibfield  {journal} {\bibinfo  {journal} {Appl. Phys. Lett.}\ }\textbf
  {\bibinfo {volume} {103}},\ \bibinfo {eid} {123109} (\bibinfo {year}
  {2013})}\BibitemShut {NoStop}%
\bibitem [{\citenamefont {Feenstra}\ \emph {et~al.}(2012)\citenamefont
  {Feenstra}, \citenamefont {Jena},\ and\ \citenamefont
  {Gu}}]{Jena-single-particle}%
  \BibitemOpen
  \bibfield  {author} {\bibinfo {author} {\bibfnamefont {R.~M.}\ \bibnamefont
  {Feenstra}}, \bibinfo {author} {\bibfnamefont {D.}~\bibnamefont {Jena}}, \
  and\ \bibinfo {author} {\bibfnamefont {G.}~\bibnamefont {Gu}},\ }\href
  {\doibase 10.1063/1.3686639} {\bibfield  {journal} {\bibinfo  {journal} {J.
  Appl. Phys.}\ }\textbf {\bibinfo {volume} {111}},\ \bibinfo {pages} {043711}
  (\bibinfo {year} {2012})}\BibitemShut {NoStop}%
\bibitem [{\citenamefont {Vasko}(2013)}]{Vasko_PRB}%
  \BibitemOpen
  \bibfield  {author} {\bibinfo {author} {\bibfnamefont {F.~T.}\ \bibnamefont
  {Vasko}},\ }\href {\doibase 10.1103/PhysRevB.87.075424} {\bibfield  {journal}
  {\bibinfo  {journal} {Phys. Rev. B}\ }\textbf {\bibinfo {volume} {87}},\
  \bibinfo {pages} {075424} (\bibinfo {year} {2013})}\BibitemShut {NoStop}%
\bibitem [{\citenamefont {Brey}(2014)}]{Brey_PRA}%
  \BibitemOpen
  \bibfield  {author} {\bibinfo {author} {\bibfnamefont {L.}~\bibnamefont
  {Brey}},\ }\href {\doibase 10.1103/PhysRevApplied.2.014003} {\bibfield
  {journal} {\bibinfo  {journal} {Phys. Rev. Applied}\ }\textbf {\bibinfo
  {volume} {2}},\ \bibinfo {pages} {014003} (\bibinfo {year}
  {2014})}\BibitemShut {NoStop}%
\bibitem [{\citenamefont {Guerrero-Becerra}\ \emph {et~al.}(2016)\citenamefont
  {Guerrero-Becerra}, \citenamefont {Tomadin},\ and\ \citenamefont
  {Polini}}]{Polini_tunneling_lifetime}%
  \BibitemOpen
  \bibfield  {author} {\bibinfo {author} {\bibfnamefont {K.~A.}\ \bibnamefont
  {Guerrero-Becerra}}, \bibinfo {author} {\bibfnamefont {A.}~\bibnamefont
  {Tomadin}}, \ and\ \bibinfo {author} {\bibfnamefont {M.}~\bibnamefont
  {Polini}},\ }\href {\doibase 10.1103/PhysRevB.93.125417} {\bibfield
  {journal} {\bibinfo  {journal} {Phys. Rev. B}\ }\textbf {\bibinfo {volume}
  {93}},\ \bibinfo {pages} {125417} (\bibinfo {year} {2016})}\BibitemShut
  {NoStop}%
\bibitem [{\citenamefont {Amorim}\ \emph {et~al.}(2016)\citenamefont {Amorim},
  \citenamefont {Ribeiro},\ and\ \citenamefont
  {Peres}}]{Phonon-assisted-theory}%
  \BibitemOpen
  \bibfield  {author} {\bibinfo {author} {\bibfnamefont {B.}~\bibnamefont
  {Amorim}}, \bibinfo {author} {\bibfnamefont {R.~M.}\ \bibnamefont {Ribeiro}},
  \ and\ \bibinfo {author} {\bibfnamefont {N.~M.~R.}\ \bibnamefont {Peres}},\
  }\href {\doibase 10.1103/PhysRevB.93.235403} {\bibfield  {journal} {\bibinfo
  {journal} {Phys. Rev. B}\ }\textbf {\bibinfo {volume} {93}},\ \bibinfo
  {pages} {235403} (\bibinfo {year} {2016})}\BibitemShut {NoStop}%
\bibitem [{\citenamefont {Ryzhii}\ \emph {et~al.}(2013)\citenamefont {Ryzhii},
  \citenamefont {Dubinov}, \citenamefont {Aleshkin}, \citenamefont {Ryzhii},\
  and\ \citenamefont {Otsuji}}]{Ryzhii_DGL_laser}%
  \BibitemOpen
  \bibfield  {author} {\bibinfo {author} {\bibfnamefont {V.}~\bibnamefont
  {Ryzhii}}, \bibinfo {author} {\bibfnamefont {A.~A.}\ \bibnamefont {Dubinov}},
  \bibinfo {author} {\bibfnamefont {V.~Y.}\ \bibnamefont {Aleshkin}}, \bibinfo
  {author} {\bibfnamefont {M.}~\bibnamefont {Ryzhii}}, \ and\ \bibinfo {author}
  {\bibfnamefont {T.}~\bibnamefont {Otsuji}},\ }\href {\doibase
  http://dx.doi.org/10.1063/1.4826113} {\bibfield  {journal} {\bibinfo
  {journal} {Appl. Phys. Lett.}\ }\textbf {\bibinfo {volume} {103}},\ \bibinfo
  {eid} {163507} (\bibinfo {year} {2013})}\BibitemShut {NoStop}%
\bibitem [{\citenamefont {Zheng}\ and\ \citenamefont
  {Das~Sarma}(1996)}]{Zheng_Lifetime}%
  \BibitemOpen
  \bibfield  {author} {\bibinfo {author} {\bibfnamefont {L.}~\bibnamefont
  {Zheng}}\ and\ \bibinfo {author} {\bibfnamefont {S.}~\bibnamefont
  {Das~Sarma}},\ }\href {\doibase 10.1103/PhysRevB.53.9964} {\bibfield
  {journal} {\bibinfo  {journal} {Phys. Rev. B}\ }\textbf {\bibinfo {volume}
  {53}},\ \bibinfo {pages} {9964} (\bibinfo {year} {1996})}\BibitemShut
  {NoStop}%
\bibitem [{\citenamefont {Hwang}\ and\ \citenamefont
  {Das~Sarma}(2007)}]{Das_Sarma_Plasmons}%
  \BibitemOpen
  \bibfield  {author} {\bibinfo {author} {\bibfnamefont {E.~H.}\ \bibnamefont
  {Hwang}}\ and\ \bibinfo {author} {\bibfnamefont {S.}~\bibnamefont
  {Das~Sarma}},\ }\href {\doibase 10.1103/PhysRevB.75.205418} {\bibfield
  {journal} {\bibinfo  {journal} {Phys. Rev. B}\ }\textbf {\bibinfo {volume}
  {75}},\ \bibinfo {pages} {205418} (\bibinfo {year} {2007})}\BibitemShut
  {NoStop}%
\bibitem [{\citenamefont {Lundeberg}\ \emph
  {et~al.}(2017{\natexlab{a}})\citenamefont {Lundeberg}, \citenamefont {Gao},
  \citenamefont {Asgari}, \citenamefont {Tan}, \citenamefont {Van~Duppen},
  \citenamefont {Alonso-Gonzalez}, \citenamefont {Woessner}, \citenamefont
  {Watanabe}, \citenamefont {Taniguchi}, \citenamefont {Hillenbrand} \emph
  {et~al.}}]{Nonlocal_GraphenePlasmons}%
  \BibitemOpen
  \bibfield  {author} {\bibinfo {author} {\bibfnamefont {M.}~\bibnamefont
  {Lundeberg}}, \bibinfo {author} {\bibfnamefont {Y.}~\bibnamefont {Gao}},
  \bibinfo {author} {\bibfnamefont {R.}~\bibnamefont {Asgari}}, \bibinfo
  {author} {\bibfnamefont {C.}~\bibnamefont {Tan}}, \bibinfo {author}
  {\bibfnamefont {B.}~\bibnamefont {Van~Duppen}}, \bibinfo {author}
  {\bibfnamefont {P.}~\bibnamefont {Alonso-Gonzalez}}, \bibinfo {author}
  {\bibfnamefont {A.}~\bibnamefont {Woessner}}, \bibinfo {author}
  {\bibfnamefont {K.}~\bibnamefont {Watanabe}}, \bibinfo {author}
  {\bibfnamefont {T.}~\bibnamefont {Taniguchi}}, \bibinfo {author}
  {\bibfnamefont {R.}~\bibnamefont {Hillenbrand}},  \emph {et~al.},\
  }\href@noop {} {\bibfield  {journal} {\bibinfo  {journal} {arXiv preprint
  arXiv:1704.05518}\ } (\bibinfo {year} {2017}{\natexlab{a}})}\BibitemShut
  {NoStop}%
\bibitem [{\citenamefont {Mishchenko}\ \emph {et~al.}(2014)\citenamefont
  {Mishchenko}, \citenamefont {Tu}, \citenamefont {Cao}, \citenamefont
  {Gorbachev}, \citenamefont {Wallbank}, \citenamefont {Greenaway},
  \citenamefont {Morozov}, \citenamefont {Morozov}, \citenamefont {Zhu},
  \citenamefont {Wong}, \citenamefont {Withers}, \citenamefont {Woods},
  \citenamefont {Kim}, \citenamefont {Watanabe}, \citenamefont {Taniguchi},
  \citenamefont {Vdovin}, \citenamefont {Makarovsky}, \citenamefont {Fromhold},
  \citenamefont {Fal'ko}, \citenamefont {Geim}, \citenamefont {Eaves},\ and\
  \citenamefont {Novoselov}}]{Twist-controlled}%
  \BibitemOpen
  \bibfield  {author} {\bibinfo {author} {\bibfnamefont {A.}~\bibnamefont
  {Mishchenko}}, \bibinfo {author} {\bibfnamefont {J.}~\bibnamefont {Tu}},
  \bibinfo {author} {\bibfnamefont {Y.}~\bibnamefont {Cao}}, \bibinfo {author}
  {\bibfnamefont {R.}~\bibnamefont {Gorbachev}}, \bibinfo {author}
  {\bibfnamefont {J.}~\bibnamefont {Wallbank}}, \bibinfo {author}
  {\bibfnamefont {M.}~\bibnamefont {Greenaway}}, \bibinfo {author}
  {\bibfnamefont {V.}~\bibnamefont {Morozov}}, \bibinfo {author} {\bibfnamefont
  {S.}~\bibnamefont {Morozov}}, \bibinfo {author} {\bibfnamefont
  {M.}~\bibnamefont {Zhu}}, \bibinfo {author} {\bibfnamefont {S.}~\bibnamefont
  {Wong}}, \bibinfo {author} {\bibfnamefont {F.}~\bibnamefont {Withers}},
  \bibinfo {author} {\bibfnamefont {C.~R.}\ \bibnamefont {Woods}}, \bibinfo
  {author} {\bibfnamefont {Y.-J.}\ \bibnamefont {Kim}}, \bibinfo {author}
  {\bibfnamefont {K.}~\bibnamefont {Watanabe}}, \bibinfo {author}
  {\bibfnamefont {T.}~\bibnamefont {Taniguchi}}, \bibinfo {author}
  {\bibfnamefont {E.~E.}\ \bibnamefont {Vdovin}}, \bibinfo {author}
  {\bibfnamefont {O.}~\bibnamefont {Makarovsky}}, \bibinfo {author}
  {\bibfnamefont {T.}~\bibnamefont {Fromhold}}, \bibinfo {author}
  {\bibfnamefont {V.}~\bibnamefont {Fal'ko}}, \bibinfo {author} {\bibfnamefont
  {A.}~\bibnamefont {Geim}}, \bibinfo {author} {\bibfnamefont {L.}~\bibnamefont
  {Eaves}}, \ and\ \bibinfo {author} {\bibfnamefont {K.}~\bibnamefont
  {Novoselov}},\ }\href@noop {} {\bibfield  {journal} {\bibinfo  {journal}
  {Nat. Nanotechnol.}\ }\textbf {\bibinfo {volume} {9}},\ \bibinfo {pages}
  {808} (\bibinfo {year} {2014})}\BibitemShut {NoStop}%
\bibitem [{\citenamefont {Lundeberg}\ \emph
  {et~al.}(2017{\natexlab{b}})\citenamefont {Lundeberg}, \citenamefont {Gao},
  \citenamefont {Woessner}, \citenamefont {Tan}, \citenamefont
  {Alonso-Gonz{\'a}lez}, \citenamefont {Watanabe}, \citenamefont {Taniguchi},
  \citenamefont {Hone}, \citenamefont {Hillenbrand},\ and\ \citenamefont
  {Koppens}}]{Koppens_thermoelectric}%
  \BibitemOpen
  \bibfield  {author} {\bibinfo {author} {\bibfnamefont {M.~B.}\ \bibnamefont
  {Lundeberg}}, \bibinfo {author} {\bibfnamefont {Y.}~\bibnamefont {Gao}},
  \bibinfo {author} {\bibfnamefont {A.}~\bibnamefont {Woessner}}, \bibinfo
  {author} {\bibfnamefont {C.}~\bibnamefont {Tan}}, \bibinfo {author}
  {\bibfnamefont {P.}~\bibnamefont {Alonso-Gonz{\'a}lez}}, \bibinfo {author}
  {\bibfnamefont {K.}~\bibnamefont {Watanabe}}, \bibinfo {author}
  {\bibfnamefont {T.}~\bibnamefont {Taniguchi}}, \bibinfo {author}
  {\bibfnamefont {J.}~\bibnamefont {Hone}}, \bibinfo {author} {\bibfnamefont
  {R.}~\bibnamefont {Hillenbrand}}, \ and\ \bibinfo {author} {\bibfnamefont
  {F.~H.}\ \bibnamefont {Koppens}},\ }\href@noop {} {\bibfield  {journal}
  {\bibinfo  {journal} {Nature materials}\ }\textbf {\bibinfo {volume} {16}},\
  \bibinfo {pages} {204} (\bibinfo {year} {2017}{\natexlab{b}})}\BibitemShut
  {NoStop}%
\bibitem [{\citenamefont {Goodfellow}\ \emph {et~al.}(2015)\citenamefont
  {Goodfellow}, \citenamefont {Chakraborty}, \citenamefont {Beams},
  \citenamefont {Novotny},\ and\ \citenamefont
  {Vamivakas}}]{Novotny_On-chip_detection}%
  \BibitemOpen
  \bibfield  {author} {\bibinfo {author} {\bibfnamefont {K.~M.}\ \bibnamefont
  {Goodfellow}}, \bibinfo {author} {\bibfnamefont {C.}~\bibnamefont
  {Chakraborty}}, \bibinfo {author} {\bibfnamefont {R.}~\bibnamefont {Beams}},
  \bibinfo {author} {\bibfnamefont {L.}~\bibnamefont {Novotny}}, \ and\
  \bibinfo {author} {\bibfnamefont {A.~N.}\ \bibnamefont {Vamivakas}},\ }\href
  {\doibase 10.1021/acs.nanolett.5b01898} {\bibfield  {journal} {\bibinfo
  {journal} {Nano Letters}\ }\textbf {\bibinfo {volume} {15}},\ \bibinfo
  {pages} {5477} (\bibinfo {year} {2015})}\BibitemShut {NoStop}%
\bibitem [{\citenamefont {Ryzhii}\ \emph {et~al.}(2014)\citenamefont {Ryzhii},
  \citenamefont {Otsuji}, \citenamefont {Aleshkin}, \citenamefont {Dubinov},
  \citenamefont {Ryzhii}, \citenamefont {Mitin},\ and\ \citenamefont
  {Shur}}]{Ryzhii_resonant_detection}%
  \BibitemOpen
  \bibfield  {author} {\bibinfo {author} {\bibfnamefont {V.}~\bibnamefont
  {Ryzhii}}, \bibinfo {author} {\bibfnamefont {T.}~\bibnamefont {Otsuji}},
  \bibinfo {author} {\bibfnamefont {V.~Y.}\ \bibnamefont {Aleshkin}}, \bibinfo
  {author} {\bibfnamefont {A.~A.}\ \bibnamefont {Dubinov}}, \bibinfo {author}
  {\bibfnamefont {M.}~\bibnamefont {Ryzhii}}, \bibinfo {author} {\bibfnamefont
  {V.}~\bibnamefont {Mitin}}, \ and\ \bibinfo {author} {\bibfnamefont {M.~S.}\
  \bibnamefont {Shur}},\ }\href {\doibase 10.1063/1.4873114} {\bibfield
  {journal} {\bibinfo  {journal} {Applied Physics Letters}\ }\textbf {\bibinfo
  {volume} {104}},\ \bibinfo {pages} {163505} (\bibinfo {year}
  {2014})}\BibitemShut {NoStop}%
\bibitem [{\citenamefont {Principi}\ \emph {et~al.}(2011)\citenamefont
  {Principi}, \citenamefont {Asgari},\ and\ \citenamefont
  {Polini}}]{Polini-soundarons}%
  \BibitemOpen
  \bibfield  {author} {\bibinfo {author} {\bibfnamefont {A.}~\bibnamefont
  {Principi}}, \bibinfo {author} {\bibfnamefont {R.}~\bibnamefont {Asgari}}, \
  and\ \bibinfo {author} {\bibfnamefont {M.}~\bibnamefont {Polini}},\ }\href
  {\doibase http://dx.doi.org/10.1016/j.ssc.2011.07.015} {\bibfield  {journal}
  {\bibinfo  {journal} {Solid State Comm.}\ }\textbf {\bibinfo {volume}
  {151}},\ \bibinfo {pages} {1627 } (\bibinfo {year} {2011})}\BibitemShut
  {NoStop}%
\bibitem [{\citenamefont {Stern}(1967)}]{Stern_2degPolarizability}%
  \BibitemOpen
  \bibfield  {author} {\bibinfo {author} {\bibfnamefont {F.}~\bibnamefont
  {Stern}},\ }\href {\doibase 10.1103/PhysRevLett.18.546} {\bibfield  {journal}
  {\bibinfo  {journal} {Phys. Rev. Lett.}\ }\textbf {\bibinfo {volume} {18}},\
  \bibinfo {pages} {546} (\bibinfo {year} {1967})}\BibitemShut {NoStop}%
\bibitem [{\citenamefont {K\"onig-Otto}\ \emph {et~al.}(2016)\citenamefont
  {K\"onig-Otto}, \citenamefont {Mittendorff}, \citenamefont {Winzer},
  \citenamefont {Kadi}, \citenamefont {Malic}, \citenamefont {Knorr},
  \citenamefont {Berger}, \citenamefont {de~Heer}, \citenamefont {Pashkin},
  \citenamefont {Schneider}, \citenamefont {Helm},\ and\ \citenamefont
  {Winnerl}}]{Malic_SlowNoncollinear}%
  \BibitemOpen
  \bibfield  {author} {\bibinfo {author} {\bibfnamefont {J.~C.}\ \bibnamefont
  {K\"onig-Otto}}, \bibinfo {author} {\bibfnamefont {M.}~\bibnamefont
  {Mittendorff}}, \bibinfo {author} {\bibfnamefont {T.}~\bibnamefont {Winzer}},
  \bibinfo {author} {\bibfnamefont {F.}~\bibnamefont {Kadi}}, \bibinfo {author}
  {\bibfnamefont {E.}~\bibnamefont {Malic}}, \bibinfo {author} {\bibfnamefont
  {A.}~\bibnamefont {Knorr}}, \bibinfo {author} {\bibfnamefont
  {C.}~\bibnamefont {Berger}}, \bibinfo {author} {\bibfnamefont {W.~A.}\
  \bibnamefont {de~Heer}}, \bibinfo {author} {\bibfnamefont {A.}~\bibnamefont
  {Pashkin}}, \bibinfo {author} {\bibfnamefont {H.}~\bibnamefont {Schneider}},
  \bibinfo {author} {\bibfnamefont {M.}~\bibnamefont {Helm}}, \ and\ \bibinfo
  {author} {\bibfnamefont {S.}~\bibnamefont {Winnerl}},\ }\href {\doibase
  10.1103/PhysRevLett.117.087401} {\bibfield  {journal} {\bibinfo  {journal}
  {Phys. Rev. Lett.}\ }\textbf {\bibinfo {volume} {117}},\ \bibinfo {pages}
  {087401} (\bibinfo {year} {2016})}\BibitemShut {NoStop}%
\bibitem [{\citenamefont {Gangadharaiah}\ \emph {et~al.}(2008)\citenamefont
  {Gangadharaiah}, \citenamefont {Farid},\ and\ \citenamefont
  {Mishchenko}}]{New_plasmon_mode}%
  \BibitemOpen
  \bibfield  {author} {\bibinfo {author} {\bibfnamefont {S.}~\bibnamefont
  {Gangadharaiah}}, \bibinfo {author} {\bibfnamefont {A.~M.}\ \bibnamefont
  {Farid}}, \ and\ \bibinfo {author} {\bibfnamefont {E.~G.}\ \bibnamefont
  {Mishchenko}},\ }\href {\doibase 10.1103/PhysRevLett.100.166802} {\bibfield
  {journal} {\bibinfo  {journal} {Phys. Rev. Lett.}\ }\textbf {\bibinfo
  {volume} {100}},\ \bibinfo {pages} {166802} (\bibinfo {year}
  {2008})}\BibitemShut {NoStop}%
\bibitem [{\citenamefont {Hwang}\ \emph {et~al.}(2007)\citenamefont {Hwang},
  \citenamefont {Hu},\ and\ \citenamefont
  {Das~Sarma}}]{Hwang_DensityDependentExchange}%
  \BibitemOpen
  \bibfield  {author} {\bibinfo {author} {\bibfnamefont {E.~H.}\ \bibnamefont
  {Hwang}}, \bibinfo {author} {\bibfnamefont {B.~Y.-K.}\ \bibnamefont {Hu}}, \
  and\ \bibinfo {author} {\bibfnamefont {S.}~\bibnamefont {Das~Sarma}},\ }\href
  {\doibase 10.1103/PhysRevLett.99.226801} {\bibfield  {journal} {\bibinfo
  {journal} {Phys. Rev. Lett.}\ }\textbf {\bibinfo {volume} {99}},\ \bibinfo
  {pages} {226801} (\bibinfo {year} {2007})}\BibitemShut {NoStop}%
\bibitem [{\citenamefont {Scott}\ \emph {et~al.}(1992)\citenamefont {Scott},
  \citenamefont {Binder},\ and\ \citenamefont {Koch}}]{Scott_Undamping}%
  \BibitemOpen
  \bibfield  {author} {\bibinfo {author} {\bibfnamefont {D.~C.}\ \bibnamefont
  {Scott}}, \bibinfo {author} {\bibfnamefont {R.}~\bibnamefont {Binder}}, \
  and\ \bibinfo {author} {\bibfnamefont {S.~W.}\ \bibnamefont {Koch}},\ }\href
  {\doibase 10.1103/PhysRevLett.69.347} {\bibfield  {journal} {\bibinfo
  {journal} {Phys. Rev. Lett.}\ }\textbf {\bibinfo {volume} {69}},\ \bibinfo
  {pages} {347} (\bibinfo {year} {1992})}\BibitemShut {NoStop}%
\bibitem [{\citenamefont {Withers}\ \emph {et~al.}(2015)\citenamefont
  {Withers}, \citenamefont {Del Pozo-Zamudio}, \citenamefont {Schwarz},
  \citenamefont {Dufferwiel}, \citenamefont {Walker}, \citenamefont {Godde},
  \citenamefont {Rooney}, \citenamefont {Gholinia}, \citenamefont {Woods},
  \citenamefont {Blake}, \citenamefont {Haigh}, \citenamefont {Watanabe},
  \citenamefont {Taniguchi}, \citenamefont {Aleiner}, \citenamefont {Geim},
  \citenamefont {Fal’ko}, \citenamefont {Tartakovskii},\ and\ \citenamefont
  {Novoselov}}]{Exciton_recombination}%
  \BibitemOpen
  \bibfield  {author} {\bibinfo {author} {\bibfnamefont {F.}~\bibnamefont
  {Withers}}, \bibinfo {author} {\bibfnamefont {O.}~\bibnamefont {Del
  Pozo-Zamudio}}, \bibinfo {author} {\bibfnamefont {S.}~\bibnamefont
  {Schwarz}}, \bibinfo {author} {\bibfnamefont {S.}~\bibnamefont {Dufferwiel}},
  \bibinfo {author} {\bibfnamefont {P.~M.}\ \bibnamefont {Walker}}, \bibinfo
  {author} {\bibfnamefont {T.}~\bibnamefont {Godde}}, \bibinfo {author}
  {\bibfnamefont {A.~P.}\ \bibnamefont {Rooney}}, \bibinfo {author}
  {\bibfnamefont {A.}~\bibnamefont {Gholinia}}, \bibinfo {author}
  {\bibfnamefont {C.~R.}\ \bibnamefont {Woods}}, \bibinfo {author}
  {\bibfnamefont {P.}~\bibnamefont {Blake}}, \bibinfo {author} {\bibfnamefont
  {S.~J.}\ \bibnamefont {Haigh}}, \bibinfo {author} {\bibfnamefont
  {K.}~\bibnamefont {Watanabe}}, \bibinfo {author} {\bibfnamefont
  {T.}~\bibnamefont {Taniguchi}}, \bibinfo {author} {\bibfnamefont {I.~L.}\
  \bibnamefont {Aleiner}}, \bibinfo {author} {\bibfnamefont {A.~K.}\
  \bibnamefont {Geim}}, \bibinfo {author} {\bibfnamefont {V.~I.}\ \bibnamefont
  {Fal’ko}}, \bibinfo {author} {\bibfnamefont {A.~I.}\ \bibnamefont
  {Tartakovskii}}, \ and\ \bibinfo {author} {\bibfnamefont {K.~S.}\
  \bibnamefont {Novoselov}},\ }\href {\doibase 10.1021/acs.nanolett.5b03740}
  {\bibfield  {journal} {\bibinfo  {journal} {Nano Letters}\ }\textbf {\bibinfo
  {volume} {15}},\ \bibinfo {pages} {8223} (\bibinfo {year} {2015})},\ \bibinfo
  {note} {pMID: 26555037},\ \Eprint
  {http://arxiv.org/abs/http://dx.doi.org/10.1021/acs.nanolett.5b03740}
  {http://dx.doi.org/10.1021/acs.nanolett.5b03740} \BibitemShut {NoStop}%
\bibitem [{\citenamefont {Khorasani}(2014)}]{Khorasani}%
  \BibitemOpen
  \bibfield  {author} {\bibinfo {author} {\bibfnamefont {S.~A.}\ \bibnamefont
  {Khorasani}},\ }\href {\doibase 10.1109/JQE.2014.2308976} {\bibfield
  {journal} {\bibinfo  {journal} {IEEE Journal of Quantum Electronics}\
  }\textbf {\bibinfo {volume} {50}},\ \bibinfo {pages} {307} (\bibinfo {year}
  {2014})}\BibitemShut {NoStop}%
\bibitem [{\citenamefont {Principi}\ \emph {et~al.}(2014)\citenamefont
  {Principi}, \citenamefont {Carrega}, \citenamefont {Lundeberg}, \citenamefont
  {Woessner}, \citenamefont {Koppens}, \citenamefont {Vignale},\ and\
  \citenamefont {Polini}}]{Principi_plasmon_loss_hBN}%
  \BibitemOpen
  \bibfield  {author} {\bibinfo {author} {\bibfnamefont {A.}~\bibnamefont
  {Principi}}, \bibinfo {author} {\bibfnamefont {M.}~\bibnamefont {Carrega}},
  \bibinfo {author} {\bibfnamefont {M.~B.}\ \bibnamefont {Lundeberg}}, \bibinfo
  {author} {\bibfnamefont {A.}~\bibnamefont {Woessner}}, \bibinfo {author}
  {\bibfnamefont {F.~H.~L.}\ \bibnamefont {Koppens}}, \bibinfo {author}
  {\bibfnamefont {G.}~\bibnamefont {Vignale}}, \ and\ \bibinfo {author}
  {\bibfnamefont {M.}~\bibnamefont {Polini}},\ }\href {\doibase
  10.1103/PhysRevB.90.165408} {\bibfield  {journal} {\bibinfo  {journal} {Phys.
  Rev. B}\ }\textbf {\bibinfo {volume} {90}},\ \bibinfo {pages} {165408}
  (\bibinfo {year} {2014})}\BibitemShut {NoStop}%
\bibitem [{\citenamefont {Vasko}\ and\ \citenamefont
  {Ryzhii}(2007)}]{Vasko-Ryzhii}%
  \BibitemOpen
  \bibfield  {author} {\bibinfo {author} {\bibfnamefont {F.~T.}\ \bibnamefont
  {Vasko}}\ and\ \bibinfo {author} {\bibfnamefont {V.}~\bibnamefont {Ryzhii}},\
  }\href {\doibase 10.1103/PhysRevB.76.233404} {\bibfield  {journal} {\bibinfo
  {journal} {Phys. Rev. B}\ }\textbf {\bibinfo {volume} {76}},\ \bibinfo
  {pages} {233404} (\bibinfo {year} {2007})}\BibitemShut {NoStop}%
\bibitem [{\citenamefont {Adam}\ \emph {et~al.}(2009)\citenamefont {Adam},
  \citenamefont {Hwang}, \citenamefont {Rossi},\ and\ \citenamefont
  {Sarma}}]{DasSarma_Impurity}%
  \BibitemOpen
  \bibfield  {author} {\bibinfo {author} {\bibfnamefont {S.}~\bibnamefont
  {Adam}}, \bibinfo {author} {\bibfnamefont {E.}~\bibnamefont {Hwang}},
  \bibinfo {author} {\bibfnamefont {E.}~\bibnamefont {Rossi}}, \ and\ \bibinfo
  {author} {\bibfnamefont {S.~D.}\ \bibnamefont {Sarma}},\ }\href {\doibase
  http://dx.doi.org/10.1016/j.ssc.2009.02.041} {\bibfield  {journal} {\bibinfo
  {journal} {Solid State Communications}\ }\textbf {\bibinfo {volume} {149}},\
  \bibinfo {pages} {1072 } (\bibinfo {year} {2009})},\ \bibinfo {note} {recent
  Progress in Graphene Studies}\BibitemShut {NoStop}%
\end{thebibliography}%

\end{document}